\documentclass[aps,amsmath,amssymb,showpacs,showkeys]{revtex4-2}

\usepackage{booktabs}
\usepackage[dvips]{graphicx}
\usepackage{times}
\usepackage{xcolor}
\usepackage{float}
\usepackage{orcidlink}
\usepackage{hyperref}
\hypersetup{
  colorlinks=true,
  urlcolor=blue,
  linkcolor=red,
  citecolor=blue
}

\newcommand{\msun}{M_{\odot}}
\newcommand{\ledd}{L_{\rm Edd}}
\newcommand{\rg}{R_g}
\newcommand{\risco}{R_{\rm ISCO}}
\newcommand{\rin}{R_{\rm in}}
\newcommand{\grs}{GRS~1739-278}
\newcommand{\fpma}{\textit{FPMA}}
\newcommand{\fpmb}{\textit{FPMB}}
\newcommand{\nustar}{\textsl{NuSTAR}}
\newcommand{\nthcomp}{\texttt{nthComp}}
\newcommand{\diskbb}{\texttt{diskbb}}
\newcommand{\gabsm}{\texttt{gabs}}
\newcommand{\tbabs}{\texttt{TBabs}}
\newcommand{\fvar}{F_{\rm var}}
\newcommand{\heasoft}{\textsc{HEASoft}}
\newcommand{\xspec}{\textsc{Xspec}}

\begin{document}

\title{\nustar\ View of the 2025 Mini-Outburst of the Black Hole X-ray Binary 
\grs: Spectral and Timing Evolution in the Soft State}

\author{Arshad Hussain\orcidlink{0009-0002-9062-2462}}
\email{arshad007h@gmail.com}

\author{Umananda Dev Goswami\orcidlink{0000-0003-0012-7549}}
\email{umananda2@gmail.com}

\affiliation{Department of Physics, Dibrugarh University, Dibrugarh 786004,
Assam, India}


\begin{abstract}
We present a dedicated timing and spectral analysis of three \nustar\ 
observations of the Galactic black hole candidate \grs\ obtained during the 
decay of its 2025 mini-outburst (September 18-30, 2025). The source is found 
in a soft, disk-dominated state throughout, with a declining $3$-$79$\,keV 
count rate (from $112.4$ to $99.4$\,cts\,s$^{-1}$), a low and only mildly 
evolving hardness ratio ($\approx0.14$-$0.16$) and a fractional variability 
that drops from $3.2\%$ to $1.1\%$; no periodic or quasi-periodic signal is 
detected in the power spectrum or the Lomb-Scargle periodogram of any 
observation. Joint \fpma+\fpmb\ spectral fits with an absorbed disk-blackbody 
plus thermal-Comptonization continuum, together with a broad, high-significance
($11.7$-$15.0\sigma$) curvature feature near $10$-$12$\,keV (an empirical 
proxy for a strong disk-reflection/returning-radiation 
component) describe the $4$-$30$\,keV spectra well 
($\chi^2/{\rm dof}=1.04$-$1.15$). The inner-disk temperature cools steadily 
from $1.021\pm0.001$ to $1.002^{+0.003}_{-0.002}$\,keV while the disk 
normalization and hence the derived inner radius, $R_{\rm in}\simeq27.7$\,km 
$\simeq0.95\,\risco$ remains constant to within $0.3\%$, consistent with a 
disk anchored at the innermost stable circular orbit throughout. The 
unabsorbed $4$-$30$\,keV luminosity declines from $1.35\times10^{37}$ to 
$1.15\times10^{37}$\,erg\,s$^{-1}$ ($0.67\%$ to $0.57\%\,\ledd$). These 
results of the spectral-timing characterization of the 2025 mini-outburst are 
broadly consistent with a high-spin black hole accreting at a low, steadily 
declining rate in the canonical soft state.
\end{abstract}


\maketitle

\section{Introduction}
\label{sec:intro}

Accreting stellar-mass black holes (BHs) in low-mass X-ray binary (LMXB) 
systems offer some of the cleanest astrophysical laboratories for studying 
matter and radiation under strong gravity. After the identification of the 
first galactic X-ray source, Sco X-1, by Giacconi et al.~\cite{giacconi62a} 
and the subsequent confirmation that Cygnus X-1 harbors a compact object which 
is too massive to be a neutron star \cite{webster72a,tananbaum72a}, more than 
twenty stellar-mass BHs have been dynamically confirmed in the Milky Way, the 
majority of them being transient LMXBs. The bulk of the emitted X-ray 
luminosity in these systems is thought to originate within a few tens of 
gravitational radii of the BH in an accretion disk, whose basic structure is 
described by the classical $\alpha$-viscosity prescription of Shakura \& 
Sunyaev \cite{shakurasunyaev73a} and its fully relativistic generalization 
was accomplished by Novikov \& Thorne \cite{novikovthorne73a}. Because the 
spacetime itself and hence the location of the innermost stable circular orbit 
(ISCO) depends on the BH spin $a$ (with gravitational radius $\rg=GM/c^2$ and 
the Kerr metric first derived by Kerr \cite{kerr63a}), the observed properties 
of the innermost disk emission - its temperature, normalization and the 
profile of any reprocessed (reflected) radiation - carry a direct imprint of 
the spin. This is the physical basis of the two principal observational 
techniques used to measure BH spin in X-ray binaries: the continuum-fitting 
method, which models the shape and normalization of the thermal disk 
continuum \cite{zhangCuiChen97a,mcclintock14a} and the relativistic reflection 
method, which models the profile of fluorescent and reflected features 
imprinted on the illuminating corona's spectrum by the strongly 
gravitationally red and blue-shifted inner disk 
\cite{fabian89a,laor91a,reynolds14a,bambi21a}.

Most LMXB BHs are transient. They spend the majority of their lives in a 
faint, quiescent state and undergo occasional outbursts, typically triggered 
by a thermal-viscous disk instability, during which the mass-accretion rate 
rises by several orders of magnitude over days to weeks before decaying back 
to quiescence over weeks to months \cite{doneGK07,tanakashibazaki96a}. During 
such an outburst, the source traces out a characteristic, hysteretic path in 
the hardness-intensity diagram (HID), moving between two principal accretion 
states separated by intermediate states \cite{homanBelloni05a,
belloniHomanCasella05a,fenderBG04a,belloni10a,mcclintockremillard06book,
vanderklis06a}. In the low/hard state, the $2$-$10$~keV spectrum is dominated 
by a Comptonized power law with photon index $\Gamma \approx 1.5$-$1.8$. Here, 
the contribution from the thermal disk is little, the source shows strong 
($10$-$30\%$ rms), broadband, red-noise variability, which is frequently 
accompanied by low-frequency (type-C) quasi-periodic oscillations (QPOs) 
\cite{casella05a} and a steady, compact radio jet \cite{fenderBG04a}. As the 
accretion rate rises, the source transits through hard and soft-intermediate 
states, in which the QPO frequency and the characteristic break frequency of 
the broadband noise increase rapidly \cite{casella05a,belloniHasinger90a}, 
into the high/soft state, in which power law steepens along with a dominant 
but weakly variable multicolor disk-blackbody component and the fractional 
rms variability typically falls below a few percent \cite{remillard06a}. In 
the soft state, the inner disk radius is believed to sit close to the ISCO, so 
that both the disk continuum and any associated relativistic reflection 
features can, in principle, be used to constrain the BH spin 
\cite{mcclintock14a,steiner10a}. At the end of the outburst, the source fades 
back through the hard state at luminosities that are typically lower than at 
the corresponding point on the rise. A hysteresis effect is usually attributed 
to a receding, advection-dominated inner flow \cite{narayan95a,maccarone03a,
kalemci13a}.

\grs\ was discovered as a bright, previously uncatalogued X-ray transient by 
the \textit{Granat} satellite in March 1996 \cite{paul96a}. A fading radio 
counterpart was subsequently identified with the Very Large Array (VLA)
telescope \cite{vargas97a}, immediately marking the source as a strong BH XRB 
candidate since radio jets are well known to accompany black holes
in their hard accretion state \cite{fenderBG04a}. A \textit{ROSAT} observation 
measured a substantial optical extinction of $A_V=14\pm2$ from the associated 
dust-scattering halo. This implied a distance of roughly $6$-$8.5$~kpc 
\cite{greiner96a}. Near-infrared photometry further showed that the companion 
star is likely a low-mass, sub-giant or later-type star, confirming that 
the \grs\ is a low-mass X-ray binary \cite{marti97a,chaty02a}. Early 
\textit{RXTE} and \textit{Mir-Kvant} monitoring of the 1996 outburst traced 
the classical hard-to-soft spectral evolution of a BH transient 
\cite{borozdin98a}. Later on, a strong, energy-dependent $\sim5$~Hz QPO was 
discovered by Borozdin \& Trudolyubov \cite{borozdin00a}. This QPO was then 
characterized in detail during the soft-intermediate state of that outburst 
by Wijnands et al.~\cite{wijnands01a}. Together, these results gave further 
support to the black hole classification and added \grs\ to the growing 
sample of transients used to study the frequency-luminosity and 
frequency-spectral-index correlations of low-frequency QPOs \cite{casella05a}.

After nearly two decades of quiescence, \grs\ returned to outburst in 
2014-2015. A \nustar\ observation taken near the peak of the rising hard 
state, at $L\approx8\%\,\ledd$, revealed a broad, relativistically skewed 
Fe\,K$\alpha$ line along with an associated reflection spectrum. This was 
analyzed by Miller et al.\ \cite{miller15a} using the relativistic reflection 
formalism developed by Fabian et al.\ and Laor \cite{fabian89a,laor91a}. That 
approach was later refined by using physically self-consistent ionized 
reflection models \cite{rossfabian05a,garciakallman10a,
garcia13a,garcia14a,dauser10a,dauser13a}. Depending on the corona geometry 
assumed, the authors inferred an accretion disk extending close to the black 
hole ($\rin=5^{+3}_{-4}\,\rg$) with a spin of $a=0.8\pm0.2$, alongside a 
compact ($\lesssim20\,\rg$) hard X-ray corona. This is broadly similar to the 
compact, variable coronae later found in other bright black hole transients
observed with \nustar\ \cite{fuerst15a,parker16a}.
A later, much fainter \nustar\ pointing, taken during the decline of the same 
outburst ($L\sim0.02\%\,\ledd$), found an unusually hard, weakly reflected 
spectrum. It also showed evidence, at the $90\%$ confidence level, for a 
truncated inner disk \cite{fuerst16a}. This kind of disk truncation is 
qualitatively similar to what has been seen in several other black hole
transients observed at very low luminosity in the hard state.

\grs\ became active again with a 2025 mini-outburst. A simultaneous 
\textit{IXPE}+\nustar\ pointing taken on September 30, 2025 - the third and 
final epoch analyzed in this paper - gave the first-ever X-ray polarimetric 
measurement of the source. Joint spectro-polarimetric modeling of this 
observation, using the \texttt{kynbbrr} returning-radiation model, favors a
near-extremal spin, $a=0.994^{+0.004}_{-0.003}$ and an inclination of 
$i=54^{+8\circ}_{-4}$. The observed polarization is best explained by 
gravitationally returning disk radiation that is reflected close to the black 
hole \cite{zhao26a}.

The 2025 mini-outburst, therefore, gives us a rare opportunity. It's a 
soft-state black hole transient, whose spin and inclination have already been 
independently pinned down through polarimetry and was observed by \nustar\ at 
more than one epoch. This lets us test something purely from the broadband 
X-ray spectral and timing behavior: does the picture painted by Zhao 
et al.\ \cite{zhao26a} for their single simultaneous observation epoch from
\nustar+\textit{IXPE} - a disk anchored at the ISCO of a near-extremal, 
moderately inclined black hole - also hold for the other, \nustar-only epochs 
of the same declining stage of the \grs? And whether the source's timing 
behavior is consistent with the weakly variable, disk-dominated corona, 
expected in this regime \cite{kara19a,uttley14a}.

In this paper, we present an independent, purely \nustar-based spectral and 
timing analysis of three observations spanning the decline of the 2025 
mini-outburst. We have four main study goals: (i) to characterize the 
broadband ($4$-$30$\,keV) continuum and any reflection-like curvature across 
the three epochs, using a joint \fpma+\fpmb\ spectral fit; (ii) to test 
whether the inner disk radius inferred from the disk continuum stays 
consistent with the ISCO throughout the decline using the mass, spin and 
inclination as reported by Zhao et al.\ \cite{zhao26a}; (iii) to characterize 
the source's aperiodic and periodic timing behavior, checking for any sign of 
state transitions or coronal activity over the 12-days span covered by these 
observations; (iv) to compare the spectral and timing evolution of this 
mini-outburst against the source's two previous \nustar-era outburst epochs 
and against the broader population of black hole X-ray binaries observed 
with \nustar.

The rest of the paper is organized as follows. Section~\ref{sec:nustar} 
briefly describes the \nustar\ observatory. Section~\ref{sec:obs} describes 
the observations and data reduction. Sections~\ref{sec:timing} and 
\ref{sec:spectral} present the timing and spectral analyses, respectively. 
Section~\ref{sec:results} discusses the results, and 
Section~\ref{sec:conclusion} summarizes our results and draws the conclusions.

\section{Nuclear Spectroscopic Telescope Array (\nustar)}
\label{sec:nustar}

\nustar\ \cite{harrison13a} is a NASA's Small Explorer mission, launched on
June 13, 2012 and remains the first and only focusing hard X-ray telescope in 
orbit. The observatory consists of two co-aligned, grazing-incidence Wolter-I 
conical optics modules mounted on opposite corners of an extendible $10$-m 
mast, focusing hard X-rays onto two nearly identical focal-plane detector 
modules, \fpma\ and \fpmb. Each of these is composed of four CdZnTe pixel 
detectors ($32\times32$ pixels, $12.3\times12.3$\,mm$^2$ per detector). 
Depth-graded Pt/C and W/Si multilayer coatings on the optics push the usable 
energy range to $3$-$79$\,keV, far beyond the $\sim10$\,keV limit of previous 
focusing X-ray telescopes, such as \textit{XMM-Newton} and \textit{Chandra}. 
\nustar does those while retaining an on-axis half-power diameter of 
$\sim18''$ and a field of view of $\sim13'\times13'$ at $10$\,keV. The energy 
resolution is $\sim400$\,eV (FWHM) at $10$\,keV and the absolute timing 
resolution is better than $2\,\mu$s. Because \nustar\ uses non-focusing CdZnTe 
detectors rather than CCDs, its data are essentially free of the photon 
pile-up that complicates the analysis of bright Galactic sources with 
CCD-based focusing telescopes. It is an important practical advantage for a 
source as bright as \grs\ during outburst.

The absolute energy and flux calibration of both focal-plane modules, together 
with their point-spread function (PSF) and effective-area models, are 
described by Madsen et al.~\cite{madsen15a}. Their report says that 
cross-calibration with other major X-ray observatories agrees to within a 
few percent over the shared bandpass. \fpma\ and \fpmb\ are physically 
separate detectors with independent electronics, which makes them useful for 
cross-checking the source spectrum and light curve against each other. They 
are conventionally fit jointly with a free multiplicative cross-normalization 
constant between the two, typically consistent with unity to within a few 
percent for a well-behaved point source \cite{harrison13a,miller15a,fuerst16a}. We follow this standard practice throughout this work, fitting \fpma\ and 
\fpmb\ jointly and reporting the \fpmb-to-\fpma\ cross-normalization constant, 
$C_{\fpmb}$, as a diagnostic of the internal calibration consistency of each 
observation.

\nustar's low-Earth orbit ($\sim600$\,km altitude, $\sim96$-minute period) has 
two important practical consequences for the present analysis. First, the 
spacecraft periodically passes through the South Atlantic Anomaly (SAA), 
during which the particle background rises sharply and data are conventionally 
removed. This, together with Earth occultations once per orbit, produces the 
characteristic gapped sampling pattern visible in our light curves 
(Section~\ref{sec:timing}) and motivates the use of gap-tolerant timing tools 
such as the Lomb-Scargle periodogram. Second, the instrumental background 
itself is non-uniform across the field of view. It includes contributions 
from unfocused (``aperture'') sky background, reflected solar X-rays, 
activation lines from cosmic-ray interactions with the spacecraft structure 
and a residual, focused cosmic X-ray background. The empirical model and 
associated \texttt{nuskybgd} software, developed by Wik et al.~\cite{wik14a}, 
were used to characterize and subtract this background. For a source as bright 
as \grs\ during the epochs studied here, with net count rates exceeding 
$\sim100$\,cts\,s$^{-1}$ per module, the background (typically 
$\lesssim0.1$\,cts\,s$^{-1}$ in the source extraction region) is a negligible 
correction. Still, we subtracted it consistently for both the spectral and 
timing products described below.

\section{Observations and Data Reduction}
\label{sec:obs}

We analyze three public \nustar\ observations of \grs\ obtained during the 
decline of its 2025 mini-outburst, spanning September 18-30, 2025 
(Table~\ref{tab:obslog}). All three were identified via the \nustar\ Master 
Catalog as pointed, non-stray-light observations of \grs\ obtained after the 
source was reported to have been re-brightened. No other public \nustar\ 
observations of the source exist between the 2014-2015 outburst epochs 
(analyzed by Miller et al.\ and F\"urst et al.\ \cite{miller15a,fuerst16a}) 
and the 2025 mini-outburst studied here. The third observation 
(ObsID 81160303004) was taken simultaneously with \textit{IXPE} and is the 
same epoch analyzed by Zhao et al.\ \cite{zhao26a}. The first two observations 
(ObsIDs 91101335002 and 81160303002) are \nustar-only pointings from the same 
decline, taken six and twelve days earlier, respectively. Neither has 
previously been the subject of a dedicated spectral-timing study.

\begin{table}[!h]
\centering
\caption{\nustar\ observation log for \grs.}
\label{tab:obslog}
\begin{tabular}{@{}lcccc@{}}
\toprule
Obs.\ & ObsID & Date & MJD & Exposure (ks) \\
\midrule
1 & 91101335002 & 2025 September 18 & 60935.0 & 15.8 \\
2 & 81160303002 & 2025 September 24 & 60941.0 & 21.7 \\
3 & 81160303004 & 2025 September 30 & 60947.0 & 21.7 \\
\bottomrule
\end{tabular}
\end{table}

Data reduction was performed with \heasoft\ v6.35.1 \cite{heasoft} and the 
corresponding \nustar\ Data Analysis Software (\texttt{NuSTARDAS} v2.1.4) 
using the \nustar\ CALDB version 20240813, the most recent calibration release 
available at the time of the analysis. For each observation and each 
focal-plane module, raw Level-1 event files were reprocessed into calibrated, 
depth-corrected and screened Level-2 event files. It was done with 
\texttt{nupipeline} by using the standard screening criteria. These include: 
the default depth cut, which rejects events that lack usable depth information 
and substantially reduces internal instrumental background at high energies;
automatic identification and removal of time intervals affected by SAA 
passages, using the \texttt{optimized} SAA method; exclusion of times when 
the source was occulted by Earth or when the pointing stability (tracked by 
the on-board star trackers) fell outside nominal bounds; and standard 
``tentacle" filtering, which rejects events associated with bright-source 
artifacts near the detector edges. No additional manual good-time-interval 
(GTI) filtering was required, as passage through the SAA and the standard 
screening already removed all strongly flaring background intervals. The 
resulting good exposure fractions were $69\%$, $73\%$ and $75\%$ of the 
on-target time for Obs.\,1-3, respectively, consistent with typical values 
for \nustar's low-Earth orbit.

Source and background products - spectra, light curves and the corresponding 
ancillary response files (ARFs, which encode the energy-dependent effective 
area) and redistribution matrix files (RMFs, which encode the detector's 
spectral redistribution) - were extracted separately for \fpma\ and \fpmb\ 
using \texttt{nuproducts}. For this, we used a circular source region of 
radius $60''$ centered on the source. This radius was chosen to capture 
$\gtrsim90\%$ of the PSF's encircled energy at $10$\,keV, while still limiting 
background and stray-light contamination. For the background, we used a 
circular region of radius $120''$ placed on a source-free part of the same 
detector, away from bad pixels, detector-gap edges or other contaminating 
point sources in the field.

All three observations have a high count rate for \grs\ 
($\gtrsim100$\,cts\,s$^{-1}$ per module; see Section~\ref{sec:timing}). So the 
background only contributes at the sub-percent level to the total count rate 
in the $4$-$30$\,keV band used for spectral fitting. We also checked that our
results don't change meaningfully under reasonable variations in the size and 
placement of the background region.

Before the timing analysis in Section~\ref{sec:timing}, we corrected the 
photon arrival times to the solar system barycenter, using the source 
coordinates and the JPL DE-200 ephemeris with \texttt{barycorr}. This step 
removes the $\sim8$-minute light-travel-time modulation caused by \nustar's 
orbital motion around the Earth-Sun barycenter. It is essential for any 
timing analysis below timescales of a few hundred seconds.

For the spectral analysis in Section~\ref{sec:spectral}, we grouped the 
extracted \fpma\ and \fpmb\ spectra of each observation to a minimum of $30$ 
counts per energy bin, using \texttt{ftgrouppha} with an optimal binning 
scheme. This grouping serves two purposes: it justifies using $\chi^2$ 
(Gaussian) statistics for the spectral fits, and it avoids the known biases 
that come from minimizing $\chi^2$ on unbinned, Poisson-distributed data with 
only a few counts per channel \cite{cash79a}.

\section{Timing Analysis}
\label{sec:timing}

\subsection{Light Curves and Hardness Ratio}

\begin{figure}[!h]
	\centerline{
	\includegraphics[scale=0.29]{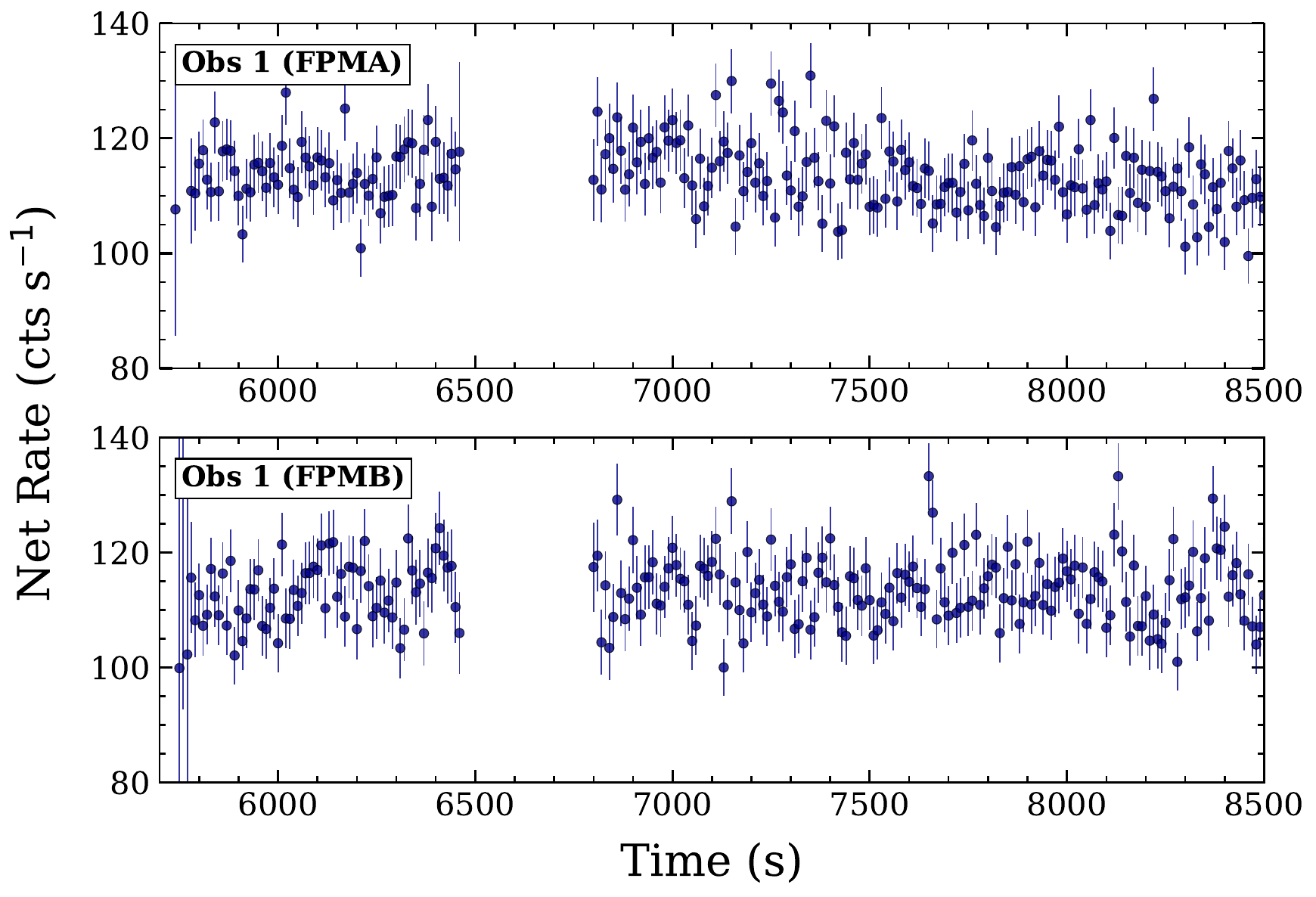}\hspace{0.5cm}
	\includegraphics[scale=0.29]{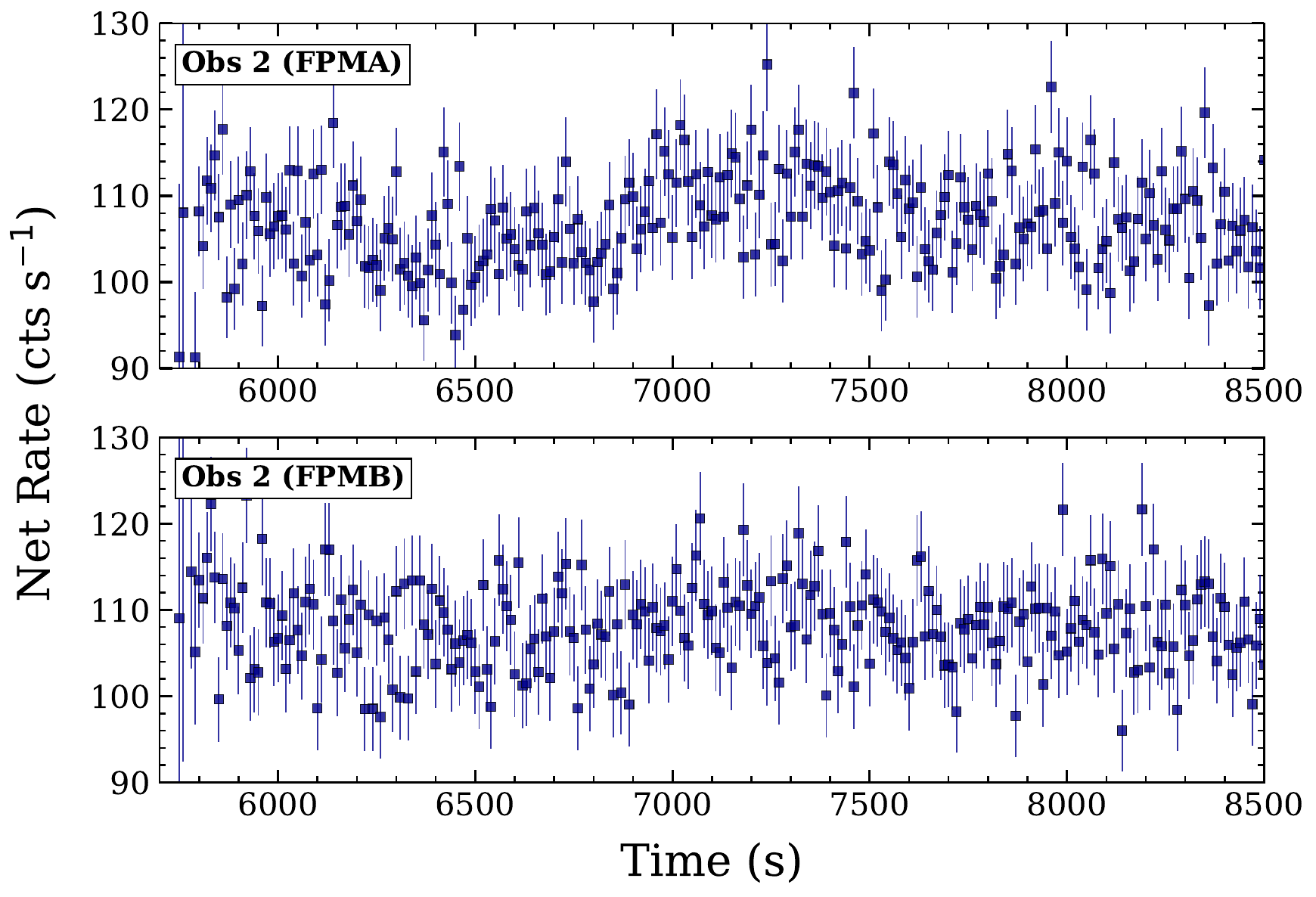}}\vspace{0.3cm}
        \centerline{
	\includegraphics[scale=0.29]{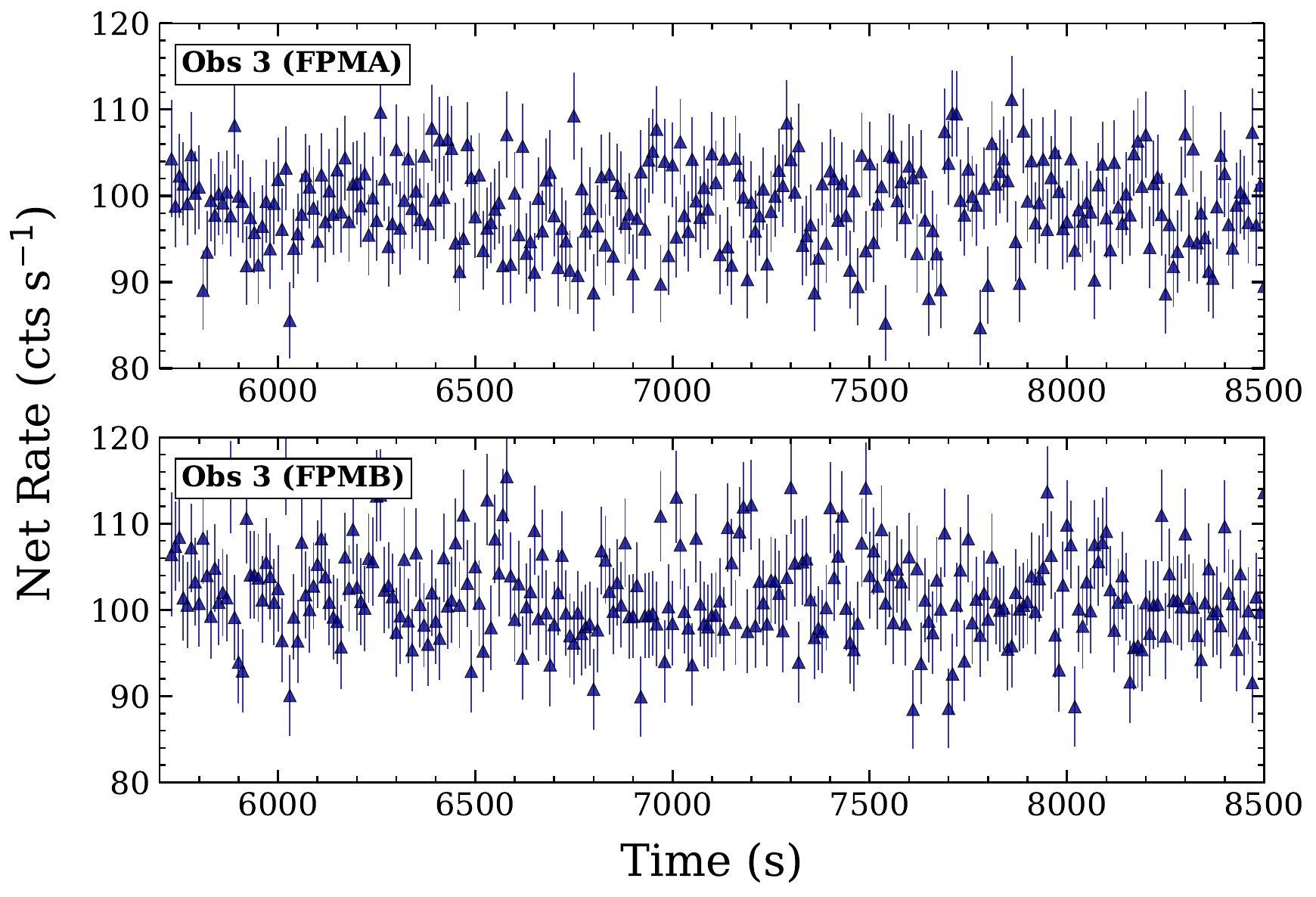}}
        \vspace{-0.1cm}
	\caption{Background-subtracted, $3\sigma$-clipped \nustar\ \fpma\ and 
\fpmb\ light curves (3-79\,keV) of \grs\ for the three 2025 mini-outburst 
observations analyzed in this work. The declining mean count rate from 
Obs.\,1 to Obs.\,3 traces the decay of the outburst.}
	\label{fig:lc}
\end{figure}

We extracted background-subtracted $3$-$79$\,keV light curves for \fpma\ and 
\fpmb\ separately, clipping at $3\sigma$ to remove bins associated with 
good-time-interval boundaries and Earth-occultation edges. This is standard 
practice for \nustar\ timing analysis of bright Galactic sources. 
Figure~\ref{fig:lc} shows the resulting light curves for all three 
observations. The mean net count rate falls steadily across the three 
epochs - from $112.4$\,cts\,s$^{-1}$ in Obs.\,1 to $108.7$\,cts\,s$^{-1}$ in 
Obs.\,2 and $99.4$\,cts\,s$^{-1}$ in Obs.\,3 (Table~\ref{tab:timing})
- tracking the overall decay of the outburst. The light curves show clear 
orbital gaps from Earth occultations and SAA passages (Section~\ref{sec:obs}), 
but no obvious flares, dips or eclipses show up within any single continuous 
data segment. This is consistent with a persistent, non-eclipsing LMXB 
geometry. This is also consistent with the fact that no orbital period shorter 
than about a day has ever been reported for \grs\, which is roughly the 
timespan covered by a single \nustar\ orbit. So there's no reason to expect 
an eclipse to show up within any one continuous stretch of our data.

\begin{figure}[!h]
	\centerline{
	\includegraphics[scale=0.28]{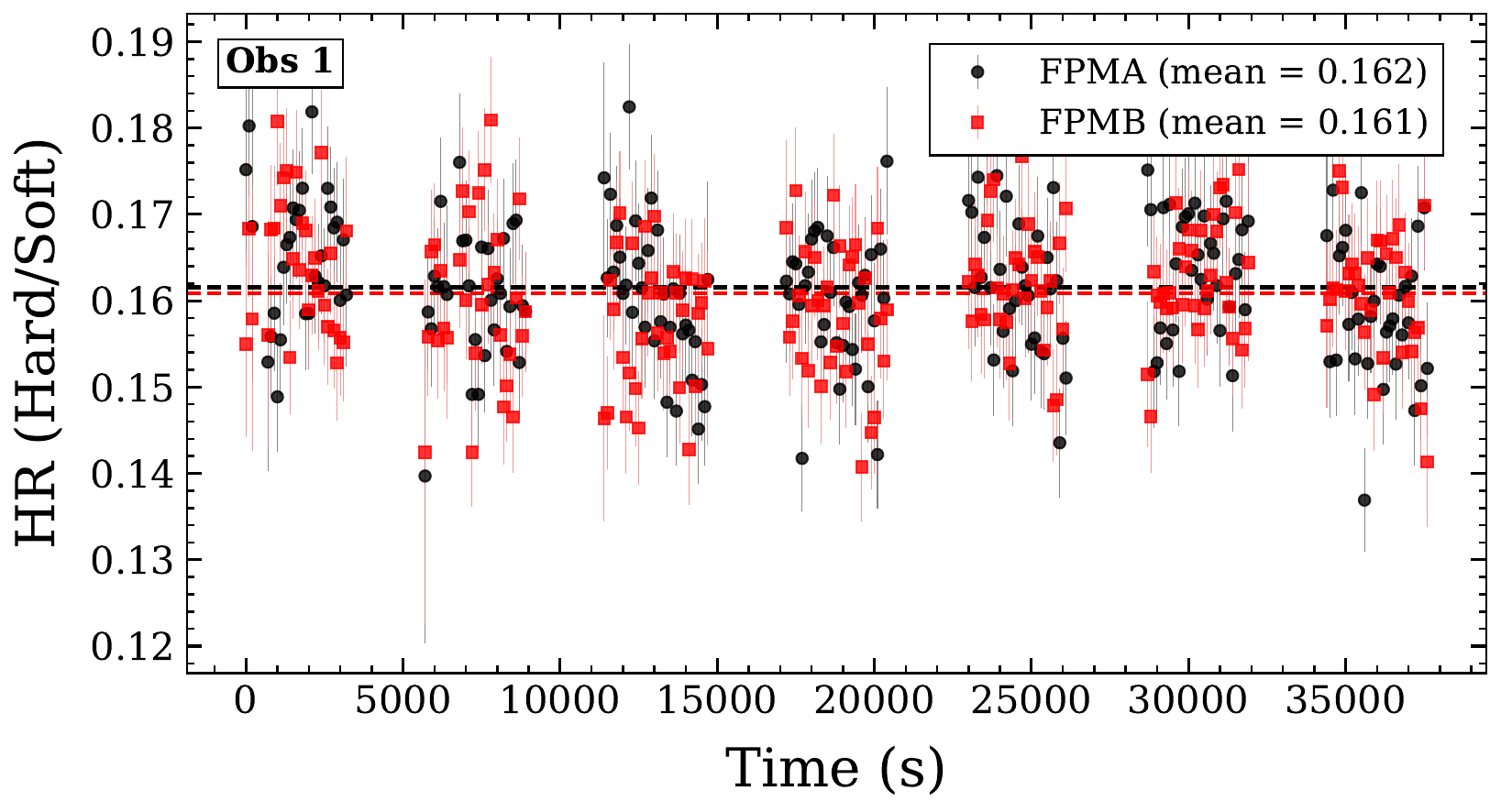} 
        \hspace{0.5cm}
	\includegraphics[scale=0.28]{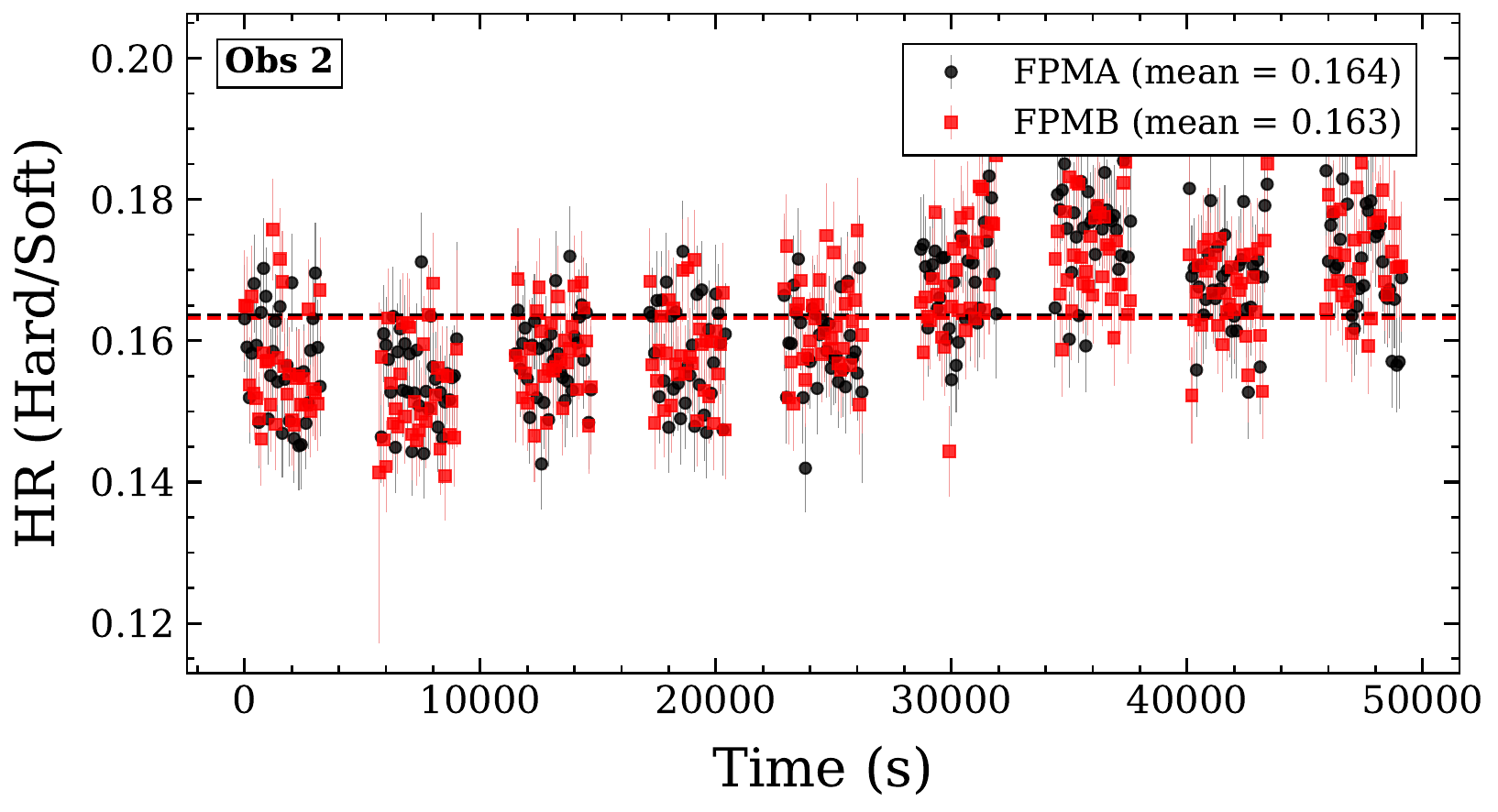}}
	\vspace{0.3cm}
        \centerline{
	\includegraphics[scale=0.28]{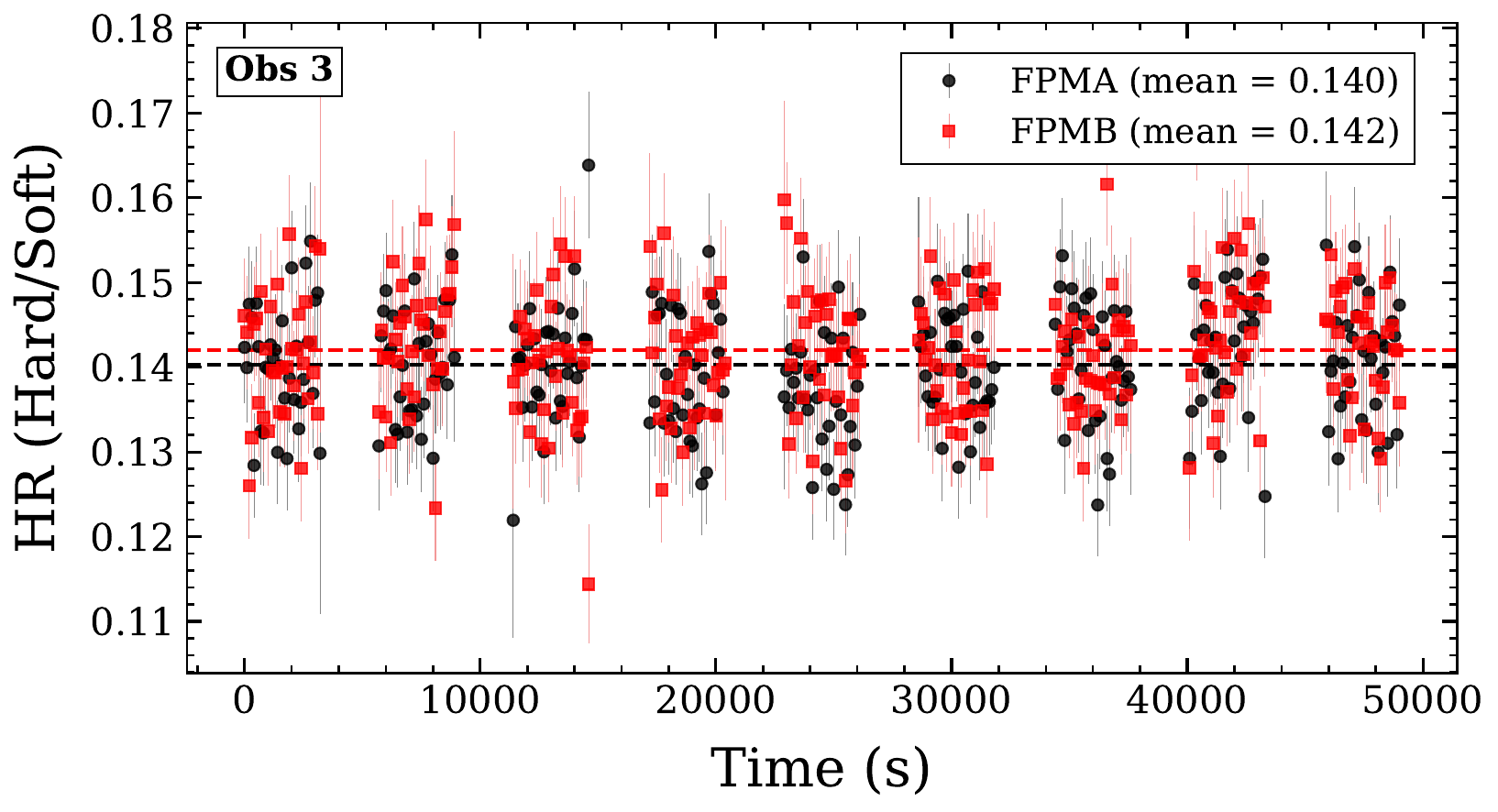}}
        \vspace{-0.1cm}
	\caption{Hardness ratio (\textit{HR} $=C_{10-79}/C_{3-10}$) light 
curves for the three observations, combining \fpma\ and \fpmb\ data.}
	\label{fig:hr}
\end{figure}

\begin{figure}[!h]
	\centerline{
	\includegraphics[scale=0.33]{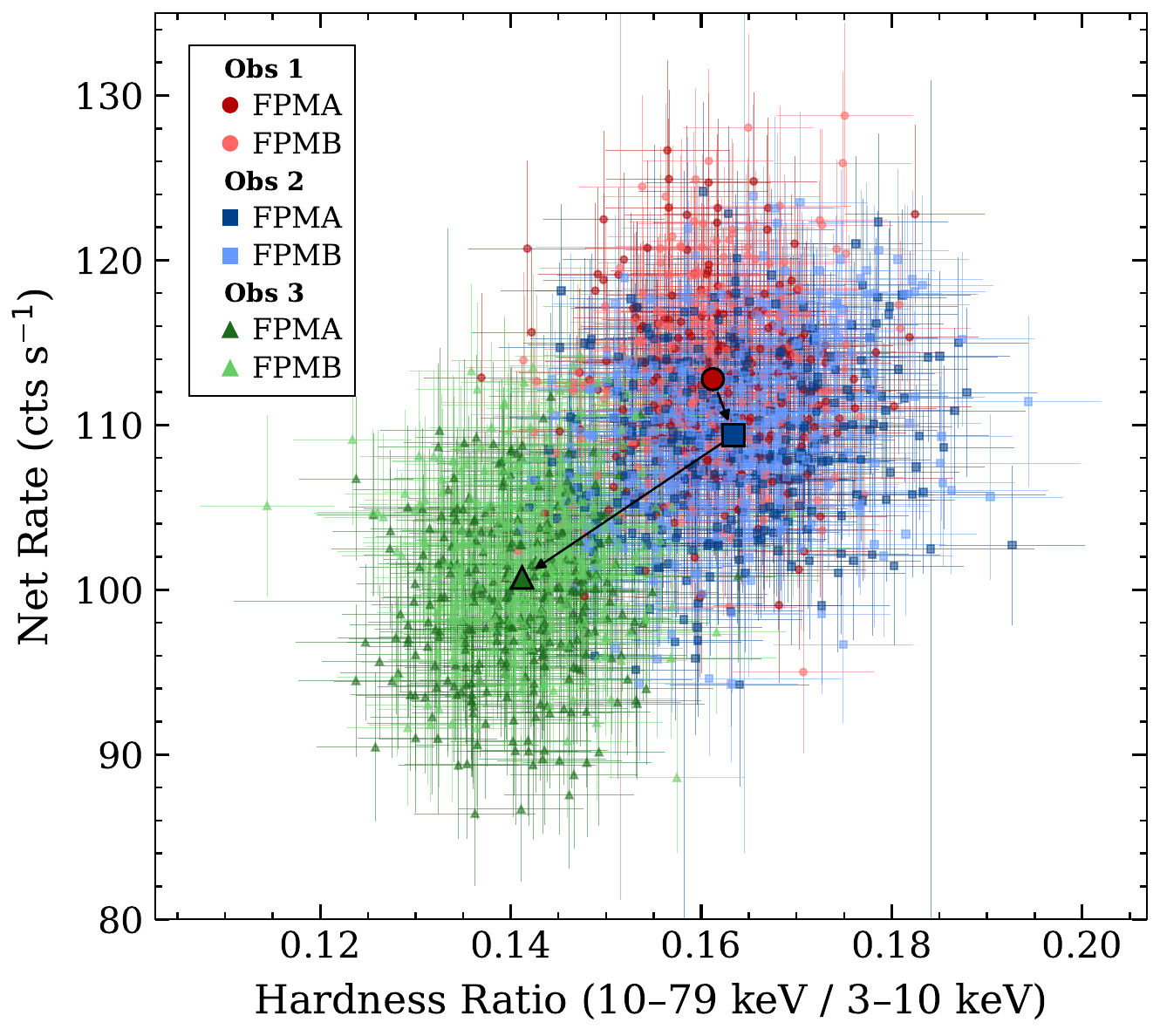}}
        \vspace{-0.2cm}
	\caption{Hardness-intensity diagram for the three 2025 mini-outburst 
\nustar\ observations of \grs. All three epochs lie on the soft branch. 
Obs.\,3 is distinctly softer and fainter than Obs.\,1 and Obs.\,2.}
	\label{fig:hid}
\end{figure}

We define a hardness ratio, \textit{HR} $\equiv C_H/C_S$, where $C_S$ and 
$C_H$ are the background-subtracted count rates in the soft ($3$-$10$\,keV) 
and hard ($10$-$79$\,keV) bands, respectively, computed in $100$ s bins. 
For a source with disk temperature $T_{\rm in}\sim1$\,keV, this 
two-band split sits roughly where the disk-blackbody emission peaks and the 
Comptonized tail begins to dominate. This 
makes it a natural way to track the relative dominance of thermal versus 
non-thermal emission in a \nustar-only, soft-state analysis like this one. 
Figure~\ref{fig:hr} shows the resulting hardness-ratio light curves and 
Figure~\ref{fig:hid} shows the corresponding hardness-intensity diagram (HID), 
which is the standard tool used to track how a black hole X-ray binary's 
spectral state evolves over an outburst. The source stays soft throughout, 
with \textit{HR} $\approx0.14$-$0.16$ in all three observations 
(Table~\ref{tab:timing}). Obs.\,1 and Obs.\,2 are nearly indistinguishable in 
the HID, while Obs.\,3 sits at a distinctly softer and fainter point, 
consistent with the outburst's continued decline. However, it is worth 
stressing that with only three epochs, we can't trace out a full HID track the 
way longer-monitored outbursts of other black hole transients follow. Our HID 
should instead be read as three snapshots taken along the soft-state decay of 
the 2025 mini-outburst.
\begin{table}[!h]
	\centering
	\caption{Counts rate (cts\,s$^{-1}$), hardness ratio (\textit{HR}) and
fractional rms variability amplitude ($\fvar$) of \grs\ (3-79\,keV) 2025 
mini-outburst observations.}
	\label{tab:timing}
	\begin{tabular}{@{}lcccccc@{}}
		\toprule
        & \multicolumn{3}{c}{\fpma} & \multicolumn{3}{c}{\fpmb} \\
	  \cmidrule(lr){2-4}   \cmidrule(lr){5-7}
		Object & Obs.\,1 & Obs.\,2 & Obs.\,3 & Obs.\,1 & Obs.\,2 
                & Obs.\,3 \\
		\midrule
		Net rate (cts\,s$^{-1}$)  & 112.4 	& 108.7   & 99.4  		& 113.0   & 109.6   & 102.4 \\
		\textit{HR} $=C_{10-79}/C_{3-10}$   & 0.162 	& 0.164   & 0.140 		& 0.161   & 0.163   & 0.142 \\
		$\fvar$ (\%)              & 3.20  	& 2.62    & 1.14  		& 2.24    & 2.34    & 1.94  \\
		\bottomrule
	\end{tabular}
\end{table}

\subsection{Fractional Variability}

\begin{figure}[!h]
	\centerline{
	\includegraphics[scale=0.3]{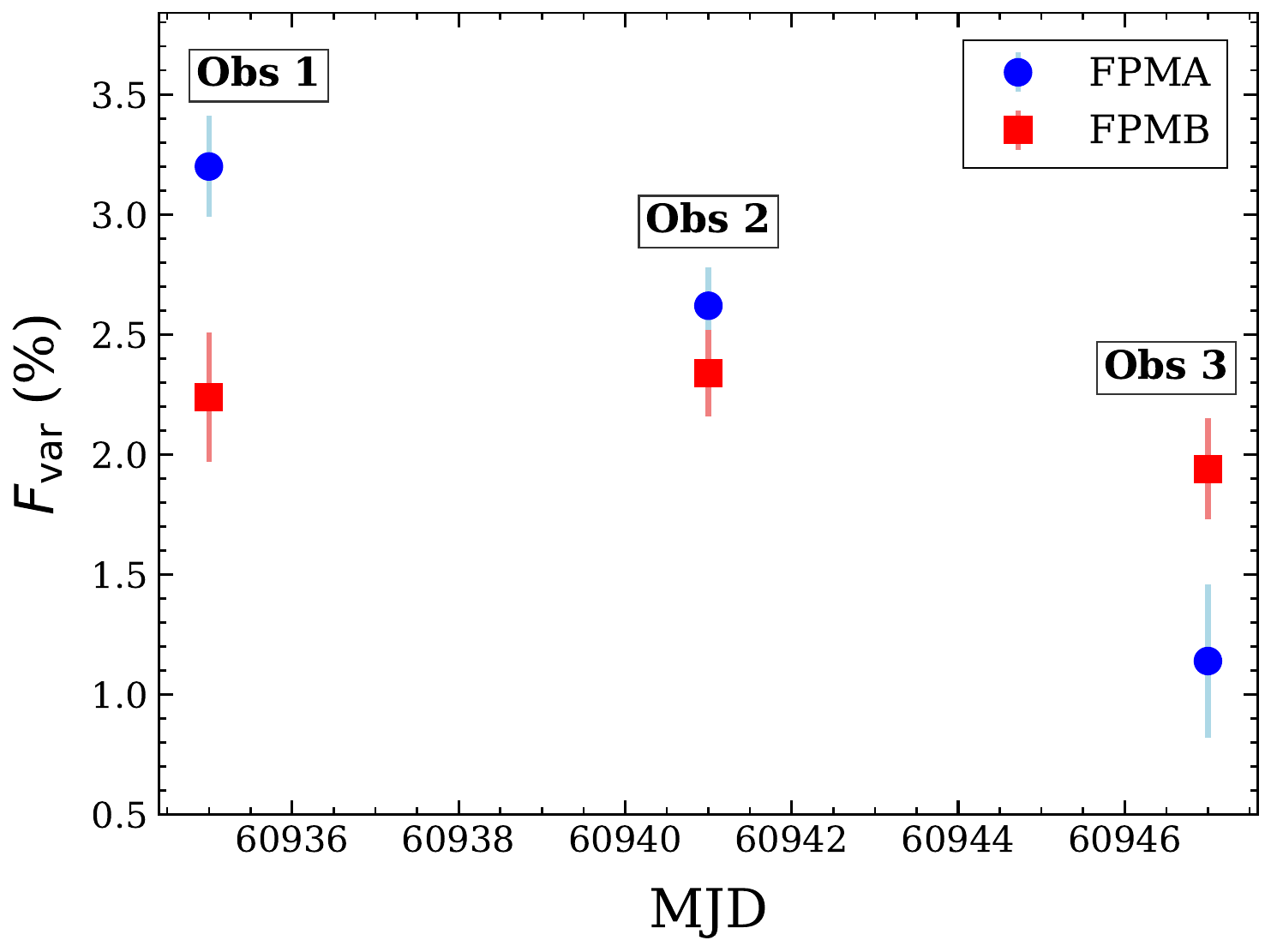}}
        \vspace{-0.1cm}
	\caption{The steady decline of the fractional rms variability amplitude 
$\fvar$ from Obs.\,1 to Obs.\,3 tracks the progressive settling of the source 
into a more strongly disk-dominated soft state.}
	\label{fig:fvar}
\end{figure}

We quantify the intrinsic aperiodic variability of the source using the 
fractional rms variability amplitude, $\fvar$, defined by Vaughan et 
al.\ \cite{vaughan03a} and computed from the $10$ s binned, 
background-subtracted, $3\sigma$-clipped $3$-$79$\,keV light curve of each 
observation as
\begin{equation}
	\fvar = \sqrt{\frac{S^2 - \overline{\sigma_{\rm err}^2}}{\bar{x}^2}},
	\label{eq:fvar}
\end{equation}
where $S^2$ is the sample variance of the $N$-point light curve, 
$\overline{\sigma_{\rm err}^2}$ is the mean square measurement (Poisson) 
error and $\bar x$ is the mean count rate. Subtracting 
$\overline{\sigma_{\rm err}^2}$ removes the Poisson counting noise, which 
would otherwise dominate, leaving behind an estimate of the intrinsic variance 
of the source itself. As summarized in Table~\ref{tab:timing} and shown in 
Figure~\ref{fig:fvar}, $\fvar$ (\fpma) decreases monotonically from $3.20\%$ 
in Obs.\,1 to $2.62\%$ in Obs.\,2 and $1.14\%$ in Obs.\,3 - roughly a factor 
of $\sim2.8$ decline over the $12$ day span of the observations. This 
behaviour matches the well-known pattern of decreasing broadband rms 
variability along the hard-to-soft evolution of black hole X-ray binary 
outbursts, where $\fvar$ typically falls from tens of percent in the hard 
state down to just a few percent (or less) deep in the soft state 
\cite{remillard06a,belloniHomanCasella05a}. It is consistent with all three 
of our observations lying on the soft branch of the outburst decline, with 
Obs.\,3 the most disk-dominated of the three. The independent \fpmb\ 
measurement shows a broadly similar decline, from $2.24\%$ in Obs.\,1 to
$1.94\%$ in Obs.\,3. However, it is less monotonic and noisier than the 
\fpma\ trend, rising slightly to $2.34\%$ in Obs.\,2 before falling in 
Obs.\,3. Therefore, we adopted \fpma\ as our reference detector for $\fvar$ 
throughout this paper. Note that the two modules agree on the overall sense 
and rough magnitude of the decline.

\subsection{RMS Spectrum}
\label{sec:rmsspec}

\begin{figure}[!h]
	\centerline{
	\includegraphics[scale=0.38]{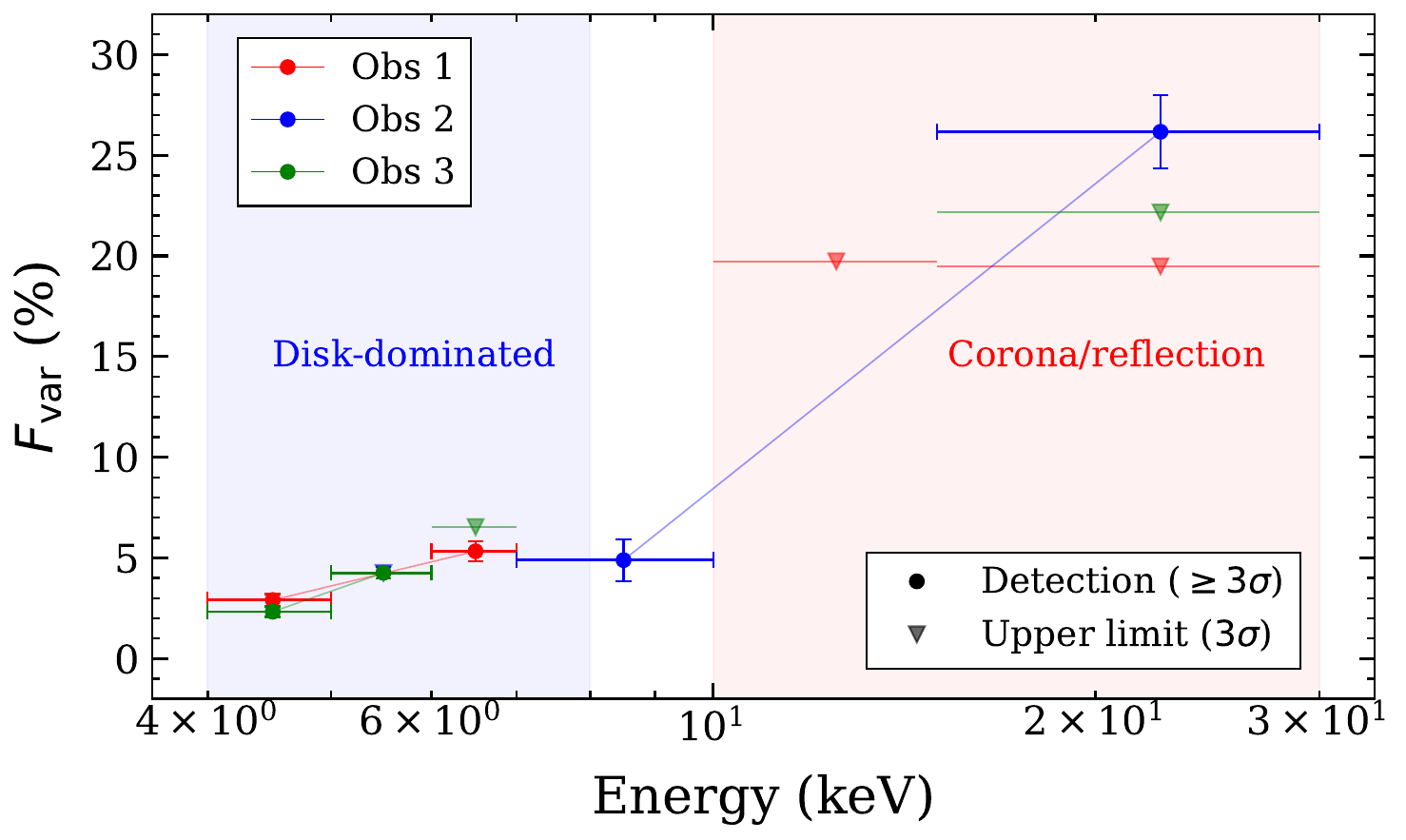}}
        \vspace{-0.1cm}
	\caption{Fractional rms (``rms spectrum'') as a function of energy 
for the three observations (\fpma), with $3\sigma$-clipped light curves in 
six energy bands spanning $4$-$30$\,keV. The comparatively flat, low-amplitude 
rms spectra in all three epochs are consistent with the low broadband 
$\fvar$ (Figure~\ref{fig:fvar}) expected for a disk-dominated soft state.}
	\label{fig:rms}
\end{figure}

Alongside the broadband $\fvar$, we can also look at how the fractional
variability changes with energy, i.e., at the so-called rms spectrum. This 
helps us identify, with more precision, which spectral component(s) is (are) 
actually responsible for the variability we observe. The technique was 
pioneered in the study of Cyg~X-1 by Revnivtsev, Gilfanov \& Churazov
\cite{revnivtsev99a}, and later generalized and reviewed by Uttley et 
al.\ \cite{uttley14a}. We computed the fractional rms amplitude in six 
adjacent energy bands spanning $4$-$30$\,keV ($4$-$5$, $5$-$6$, $6$-$7$, 
$7$-$10$, $10$-$15$ and $15$-$30$\,keV) for each observation, using \fpma\ 
light curves with $100$ s time bins. We applied the same $3\sigma$-clipping 
and Poisson-noise subtraction used for the broadband $F_{\rm var}$ 
(Equation~\ref{eq:fvar}). Figure~\ref{fig:rms} shows the resulting rms spectra 
for the three observations. We only show energy bands where the detection 
exceeds $3\sigma$ significance. Bins that fall below this threshold are 
instead treated as upper limits.

In a standard truncated-disk/hot-flow or disk-plus-corona picture, an rms 
spectrum that rises with energy indicates that the variability is dominated 
by the Comptonized component while the disk emission is comparatively steady. 
On the other hand, a flat or declining rms spectrum indicates that disk and 
coronal variability are more closely coupled or that the disk itself 
contributes appreciably to the observed variability \cite{uttley14a,
revnivtsev99a}. We find that the rms spectrum of \grs\ is comparatively flat 
and low-amplitude ($\lesssim5\%$) across the full \nustar\ bandpass in all 
three observations, with a mild rise toward the hardest energies. It is 
consistent with the low broadband $\fvar$ measured directly from the light 
curves. This also fits with a soft-state configuration in which both the disk 
and the comparatively faint, non-dominant Comptonized tail contribute little 
intrinsic variability over the whole observations. Because our rms spectrum 
is computed over each observation's full multi-hour exposure, it only captures 
variability on timescales of minutes to roughly an orbit. It therefore cannot 
probe the much faster, sub-second reverberation signals that studies like 
Kara et al.~and Uttley et al.~\cite{kara19a,uttley14a} use to measure how 
compact a corona is in brighter, harder-state sources.

\subsection{Power Spectrum, Cross-Correlation, and Periodicity Search}

We searched for broadband ($0.5\%$-$10\%$ rms level) noise and for periodic or 
quasi-periodic signals using the open-source spectral-timing package 
\texttt{Stingray}, developed by Huppenkothen et al.\ \cite{huppenkothen19a}. 
Power spectra were built using the standard ``fractional-rms" scaling 
(Miyamoto normalization), which puts the variability in physically meaningful 
units rather than raw counts. We then removed the Poisson noise contribution 
using a data-driven method, following the general approach reviewed by Uttley 
et al.\ \cite{uttley14a}. This overall technique traces back to earlier 
foundational work by Lyubarskii \cite{lyubarskii97a} and others on modeling 
variability in black hole X-ray binaries. In none of the three observations do 
we find a broadband noise component or a discrete QPO feature above the noise 
floor at a statistically significant level. This null result is consistent 
with the low $\fvar$ and flat rms spectrum measured in all three epochs and 
with the expected weak-corona in a disk-dominated soft state. It is worth 
noting that this is quite different from the strong, evolving low-frequency 
QPOs typically seen in the hard and intermediate states of black hole 
transients, including the $\sim5$\,Hz QPO previously detected in \grs\ itself 
during the 1996 outburst \cite{borozdin00a,wijnands01a}. These QPOs are 
usually explained as Lense-Thirring precession of a geometrically thick, 
truncated inner flow \cite{ingram09a}. That is a geometry that, by definition, 
shouldn't apply to the disk-dominated soft-state epochs we studied here.

We also cross-correlated the soft ($3$-$10$\,keV) and hard ($10$-$79$\,keV) 
band light curves using the discrete correlation function (DCF), simply to 
check whether hard-band photons lag behind soft-band photons in time. It is a 
pattern that could point to reverberation or coupling between the corona and 
the disk. This technique was first used this way on Cyg~X-1 by Miyamoto \& 
Kitamoto \cite{miyamotokitamoto89a}. Later on, this was refined statistically 
by Revnivtsev, Gilfanov \& Churazov \cite{revnivtsev99a}. The DCF amplitude is 
consistent with zero across all tested lags ($\pm2000$\,s) in Obs.\,1 and 
Obs.\,3. Obs.\,2 shows a weak, broadband elevation ($\mathrm{DCF}\sim0.2$) 
without a well-defined peak. It may be connected to the somewhat stronger 
hard-band variability in that observation. But this does not correspond to 
a statistically significant lag detection. 

Finally, we searched for periodic signals using the Lomb-Scargle periodogram 
\cite{lomb76a,scargle82a}, which is appropriate for the unevenly sampled, 
gapped light curves resulting from \nustar's low-Earth orbit. No peak exceeds 
the $1\%$ false-alarm-probability threshold in any observation. We therefore 
report a formal non-detection of periodic or quasi-periodic variability in 
all three epochs.

\section{Spectral Analysis}
\label{sec:spectral}

Spectral fitting was performed jointly for \fpma\ and \fpmb\ in the 
$4.0$-$30.0$\,keV band using the \xspec\ fitting package described by Arnaud 
\cite{arnaud96a} (v12.15.0). 
Interstellar absorption was modeled with \tbabs, the ISM absorption 
model of Wilms et al.\ \cite{wilms00a}. It computes the wavelength-dependent 
opacity of the neutral gas and dust along the line of sight from the column 
density $N_{\rm H}$, for a given elemental abundance pattern and set of 
photoionization cross-sections.
Interstellar abundances were set to the values of 
Wilms et al.\ \cite{wilms00a} and photoelectric absorption cross-sections were 
set to those of Verner et al.\ \cite{verner96a}. Both are now standard choices 
for X-ray binary spectral analysis and replace the older abundance tables that 
tend to underestimate $N_{\rm H}$ for a given absorption depth.  

We restricted our fits to $4.0$-$30.0$\,keV rather than the full \nustar\ 
bandpass for two reasons. Below $\sim4$\,keV, the effective area and 
background modeling become increasingly uncertain for a source of this 
brightness. Above $\sim30$\,keV, the already faint Comptonized tail of \grs\ 
in its soft state spectrum contributes little statistical weight, while 
adding correlated systematic uncertainty from residual instrumental background.
Throughout, we allow a free multiplicative cross-normalization constant 
$C_{\fpmb}$ between the two focal-plane modules (\fpma\ fixed at unity). 
Every quoted parameter uncertainty here is at the $90\%$ confidence level 
($\Delta\chi^2 = 2.706$ for one parameter at a time) unless stated otherwise.

\begin{figure}[!h]
        \centerline{
        \includegraphics[scale=0.3]{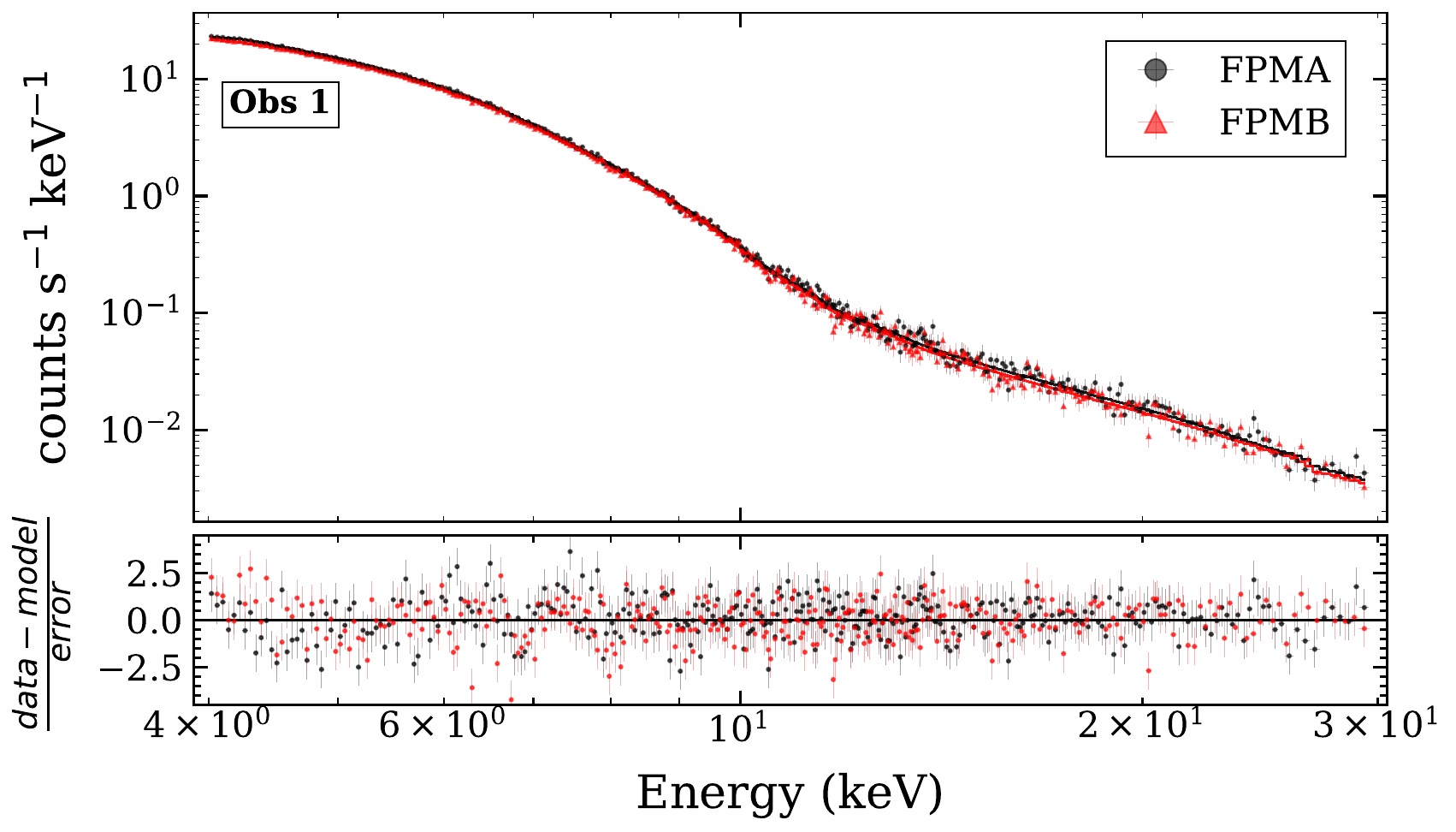}
        \hspace{0.3cm}
        \includegraphics[scale=0.3]{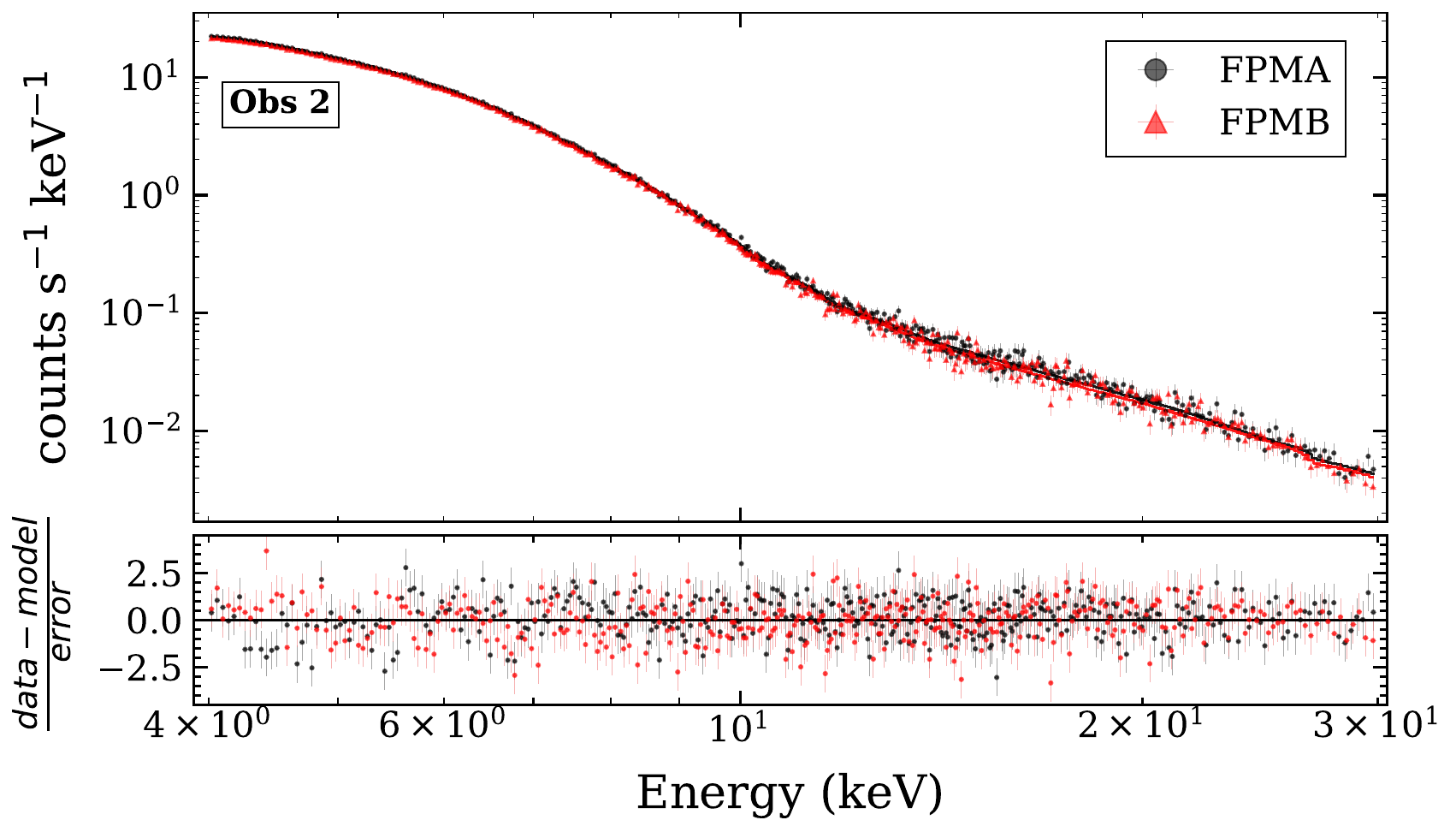}}
        \vspace{0.3cm}
        \centerline{
        \includegraphics[scale=0.3]{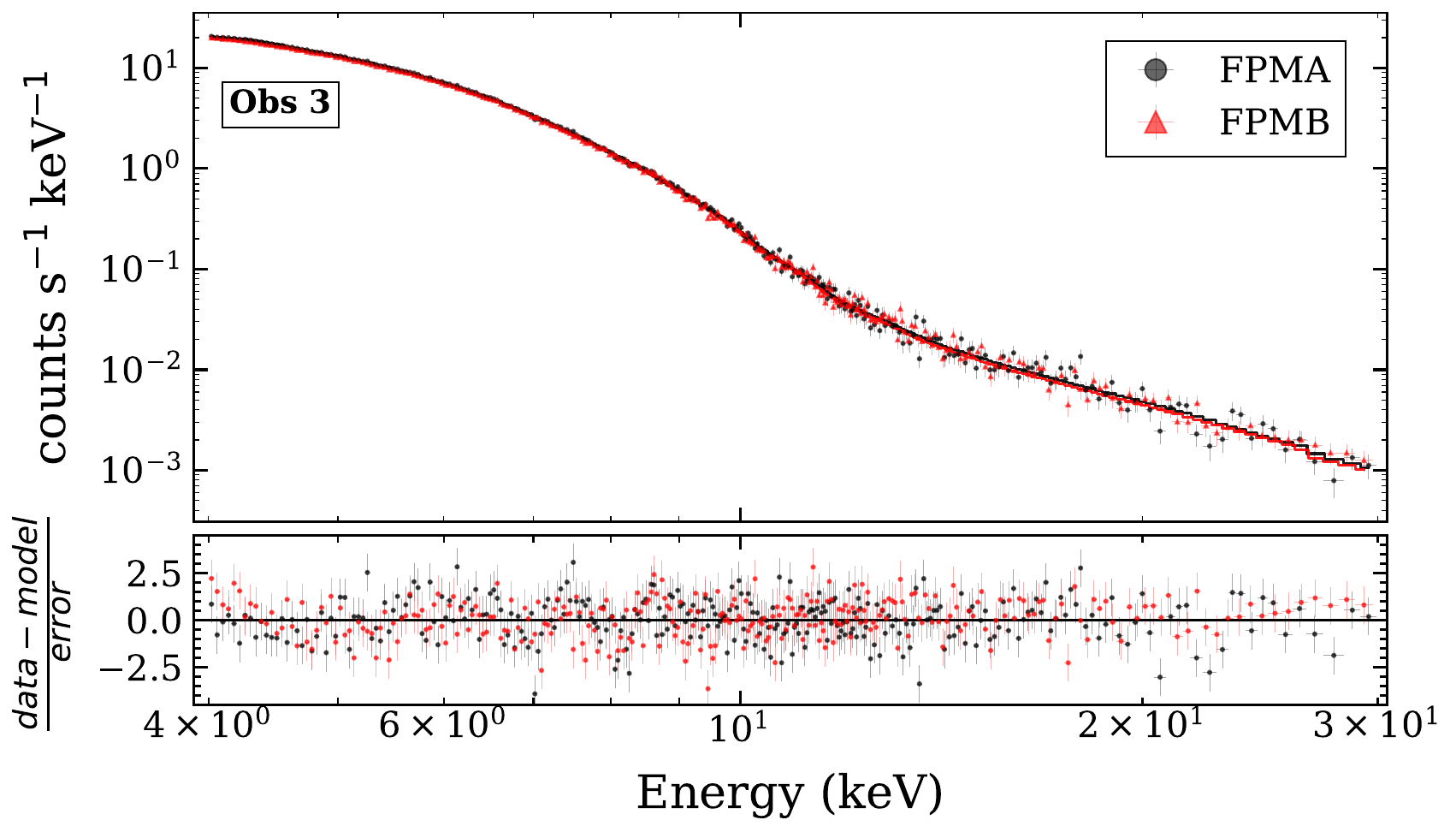}}
        \vspace{-0.1cm}
        \caption{Joint \fpma\ (black) + \fpmb\ (red) spectral fits to the
three \grs\ observations with the model
$C_{\rm det}\times\tbabs\times\gabsm\times(\diskbb+\nthcomp)$ and residuals
in units of $\sigma$.}
        \label{fig:spec}
\end{figure}
We began with a simple absorbed multicolor disk-blackbody, following the model 
of Mitsuda et al.\ \cite{mitsuda84a}, plus power-law continuum, 
$\tbabs\times(\diskbb+\texttt{powerlaw})$. This baseline fit is statistically 
unacceptable in all three observations ($\chi^2/{\rm dof}\gtrsim1.3$). It 
gave broad, systematic residuals in the $8$-$15$\,keV range. This resembles 
the kind of residual structure that has led to the adoption of relativistic 
reflection models across the BH XRB population \cite{fabian89a,rossfabian05a,
garciakallman10a}.

We first tried adding a relativistically broadened Fe\,K$\alpha$ line, using 
the \texttt{laor} model of Laor \cite{laor91a} with the line's rest-frame 
energy fixed at $6.4$\,keV. This did not improve the fit by a statistically 
meaningful amount in any of the three observations. In other words, our data 
don't favor a relativistically blurred iron line as the origin of the 
curvature we see. Without soft X-ray coverage below $\sim3$\,keV to 
independently constrain the continuum and absorption column, a real 
relativistically broadened Fe line is difficult to distinguish from other 
broad curvatures in the spectrum. \nustar\ data alone, spanning only 
$3$-$79$\,keV, can't fully break that degeneracy.
 
A fully self-consistent reflection model like \texttt{relxill} 
\cite{garcia14a,dauser13a} or \texttt{reflionx} \cite{rossfabian05a} would 
in principle be the more physical choice. However, these models could not 
accommodate the corona geometry, ionization state and iron abundance features 
in a single fit at once. With \nustar\ data alone in this faint, 
soft-state regime, those parameters simply aren't well constrained. So instead 
we took a simpler, purely empirical approach, we added a broad Gaussian 
absorption-like component (\gabsm) around $10$-$12$\,keV. We are not claiming 
that this represents a real absorption feature. We are just using it as a 
flexible, model-independent way to soak up the broad curvature that a 
genuine disk-reflection or returning-radiation component would leave in the 
$4$-$30$\,keV NuSTAR band. This component produced a large and highly 
significant improvement in all three fits (Section~\ref{sec:ftest}) and is 
retained in the final model. 

\begin{figure}[!h]
	\centerline{
	\includegraphics[scale=0.32]{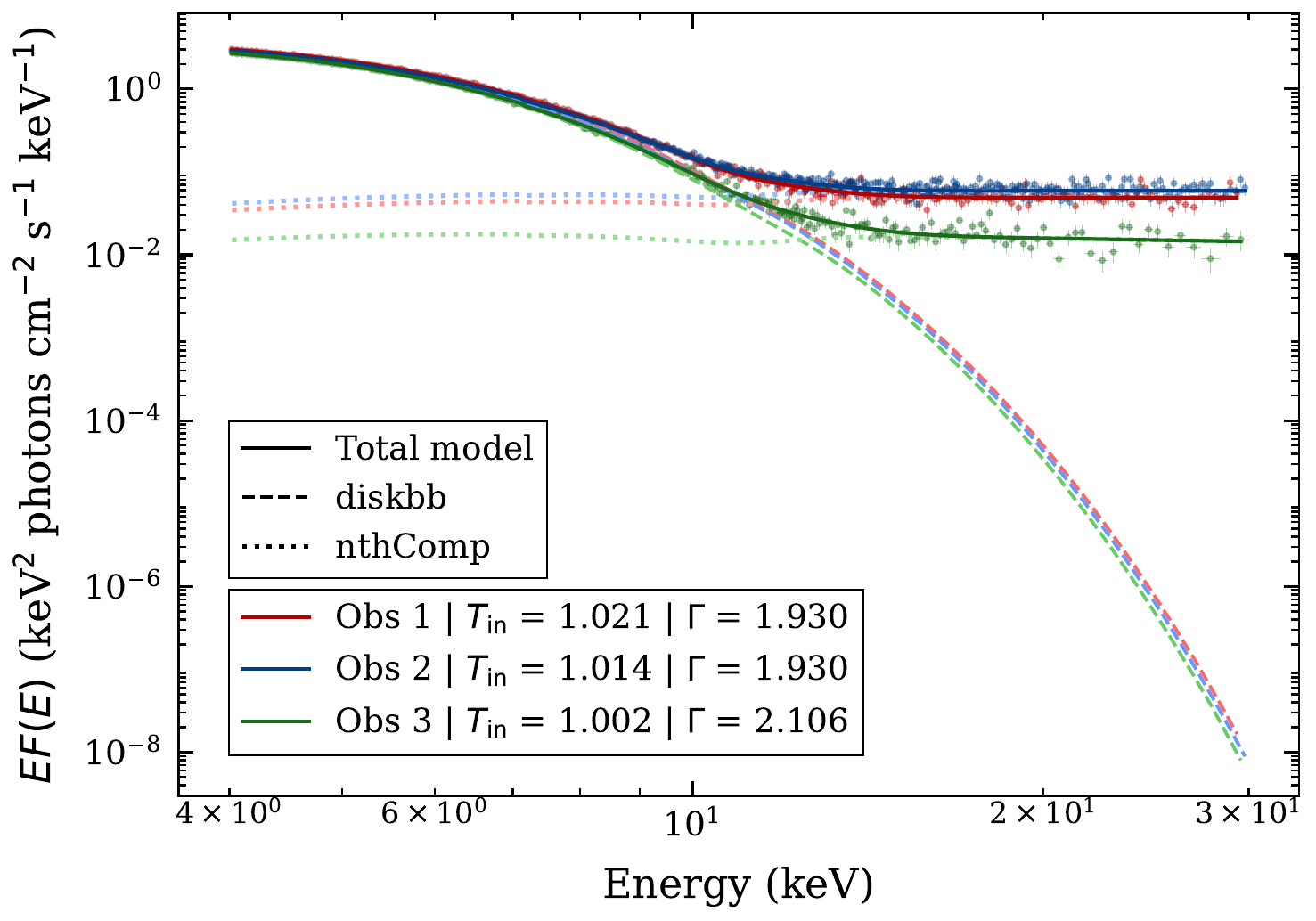}}
        \vspace{-0.1cm}
	\caption{Unfolded $EF(E)$ spectra for the three observations (\fpma), 
with \diskbb\ (dashed) and \nthcomp\ (dotted) components shown separately 
from the total model (solid).}
	\label{fig:eeuf}
\end{figure}

Finally, we swapped out the simple power-law for a more physical model called 
thermal Comptonization, using the \nthcomp\ model of Zdziarski, Johnson \& 
Magdziarz and \.Zycki, Done \& Smith \cite{zdziarski96a,zycki99a}. It is 
itself built on the Comptonization framework developed by Sunyaev \& 
Titarchuk \cite{sunyaev80a}. We tied the seed-photon temperature 
$kT_{\rm bb}$ to the disk temperature $T_{\rm in}$, since the seed photons 
for Comptonization should physically come from the same disk. We froze the 
electron temperature at $kT_e=50$\,keV, since our $4$-$30$\,keV \nustar\ data 
can't constrain it on their own for a spectrum this soft and only weakly 
Comptonized. The main advantage of \nthcomp\ over the plain power-law is that 
it doesn't have the power-law's unphysical problem of extending to 
indefinitely high energies. It turns over at high energy in a physically 
sensible way. The interstellar column was frozen at 
$N_{\rm H}=3.91\times10^{22}$\,cm$^{-2}$ in all fits to avoid a known 
degeneracy between $N_{\rm H}$ and the broad \gabsm\ component given 
\nustar's limited sensitivity below $\sim4$\,keV. Thus, the adopted final 
model is
\begin{equation}
	C_{\rm det}\times \tbabs(N_{\rm H})\times\gabsm\times\big[\diskbb(T_{\rm in})+\nthcomp(\Gamma,kT_e,kT_{\rm bb})\big],
\end{equation}
It is fitted jointly to \fpma\ and \fpmb\ data for each of the three 
observations, with all \fpmb\ model parameters (other than $C_{\fpmb}$) 
tied to the corresponding \fpma\ values. This model provides a good description of all three spectra, with $\chi^2/{\rm dof}=1.101$, 
$1.037$ and $1.151$ for Obs.\,1-3, respectively (Table~\ref{tab:specpar}). 
Figure~\ref{fig:spec} shows the data, best-fit model and residuals for all 
three epochs. Figure~\ref{fig:eeuf} shows the corresponding unfolded $EF(E)$ 
spectra, with the disk and Comptonization components shown separately. 
In this representation, the model's photon flux $F(E)$ is multiplied 
by energy $E$, so that the area under the curve on a logarithmic plot traces 
the source's energy output directly, making it easy to see at a glance which 
component and which part of the spectrum dominates the radiated power.

\begin{table}[!h]
	\centering
	\caption{Best-fit joint \fpma+\fpmb\ spectral parameters for the model 
$C_{\rm det}\times\tbabs(N_{\rm H})\times\gabsm\times(\diskbb+\nthcomp)$ within
$4$-$30$\,keV energy range. Errors are at the $90\%$ confidence level ($\Delta\chi^2=2.706$).}
	\label{tab:specpar}
	\begin{tabular}{@{}lccc@{}}
\toprule
Parameter & Obs.\,1 & Obs.\,2  & Obs.\,3 \\
\midrule
$N_{\rm H}$ ($10^{22}$\,cm$^{-2}$) & $3.91$ (frozen) & $3.91$ (frozen) & $3.91$ (frozen) \\
$C_{\fpmb}$ & $1.002\pm0.003$ & $1.010\pm0.003$ & $1.027^{+0.003}_{-0.002}$ \\
gabs $E_{\rm line}$ (keV) & $10.639^{+0.235}_{-0.216}$  & $10.427^{+0.181}_{-0.172}$ & $10.828^{+0.286}_{-0.240}$  \\
gabs $\sigma$ (keV) & $1.146^{+0.180}_{-0.165}$ & $1.067^{+0.147}_{-0.133}$     & $1.426^{+0.207}_{-0.183}$ \\
gabs Strength & $0.438^{+0.118}_{-0.095}$ & $0.335^{+0.074}_{-0.063}$ & $0.772^{+0.216}_{-0.158}$ \\
$T_{\rm in}$ (keV) & $1.021\pm0.001$ & $1.014\pm0.002$ & $1.002^{+0.003}_{-0.002}$ \\
$N_{\diskbb}$ & $650.5\pm8.9$ & $650.8\pm7.8$ & $652.5^{+9.7}_{-10.4}$ \\
$\Gamma$ & $1.930^{+0.077}_{-0.072}$ & $1.930^{+0.053}_{-0.050}$ & $2.106^{+0.151}_{-0.135}$ \\
$kT_e$ (keV) & $50$ (frozen) & $50$ (frozen) & $50$ (frozen) \\
$N_{\nthcomp}$ ($10^{-2}$) & $1.421^{+0.24}_{-0.20}$ & $1.723^{+0.19}_{-0.17}$  & $0.678^{+0.23}_{-0.16}$ \\
$\chi^2$/dof & $766.11/696$ & $841.63/812$ & $666.84/579$ \\
$\chi^2_\nu$ & $1.101$ & $1.037$  & $1.151$ \\
\bottomrule
\end{tabular}
\end{table}

\subsection{Significance of the Reflection-like Curvature}
\label{sec:ftest}

We assessed the statistical significance of the \gabsm\ component 
using the standard $\chi^2$ $F$-test described by Bevington \& Robinson 
\cite{bevington03a}. We compared the best-fit $\chi^2$ with and without the 
component (the latter obtained by freezing the \gabsm\ strength at zero and 
refitting). There is one important statistical caution applied here, as 
discussed by Protassov et al.~\cite{protassov02a}. When testing for a new 
component like \gabsm\ feature, the standard $F$-test does not behave in a 
perfectly reliable way. This is because the null hypothesis places the 
relevant parameters (centroid and width) at the boundary of the parameter 
space. Here, we do not treat our quoted significances - $11.7\sigma$, 
$12.4\sigma$ and $15.0\sigma$ - as rigorously calibrated detection 
probabilities. Instead, we treat them simply as a sign that the data strongly 
prefer including the extra component over leaving it out. A fully rigorous 
calibration would require Bayesian methods or simulation-based calibration, 
which is beyond the scope of this \nustar-only study. Even with this 
limitation in mind, the evidence for the feature is strong. As 
Table~\ref{tab:ftest} shows, adding the \gabsm\ component improves the fit 
by a large amount in all three observations: $\Delta\chi^2=166.7$, $176.0$ 
and $316.5$, for just one extra free parameter. Even if a more careful, 
conservative calibration may bring these numbers down substantially, the 
improvement is still large enough in all three observations. This still 
supports our main conclusion that there is really a strong, broad curvature 
feature in the $10$-$12$\,keV range.
\begin{table}[!h]
\centering
\caption{$F$-test results for the \gabsm\ component (joint \fpma+\fpmb).}
\label{tab:ftest}
\begin{tabular}{@{}lccccc@{}}
\toprule
Obs. & $\chi^2/{\rm dof}$ (with gabs) & $\chi^2/{\rm dof}$ (w/o gabs) & $\Delta\chi^2$ & $F$-statistic & Significance \\
\midrule
~~~1   & $766.11/696$ & $932.84/697$ & $166.7$ & $151.5$ & $11.7\sigma$ \\
~~~2   & $841.63/812$ & $1017.63/813$ & $176.0$ & $169.8$ & $12.4\sigma$ \\
~~~3   & $666.84/579$ & $983.34/580$ & $316.5$ & $274.8$ & $15.0\sigma$ \\
\bottomrule
\end{tabular}
\end{table}

\subsection{Flux and Luminosity}

\begin{figure}
	\centerline{
	\includegraphics[scale=0.3]{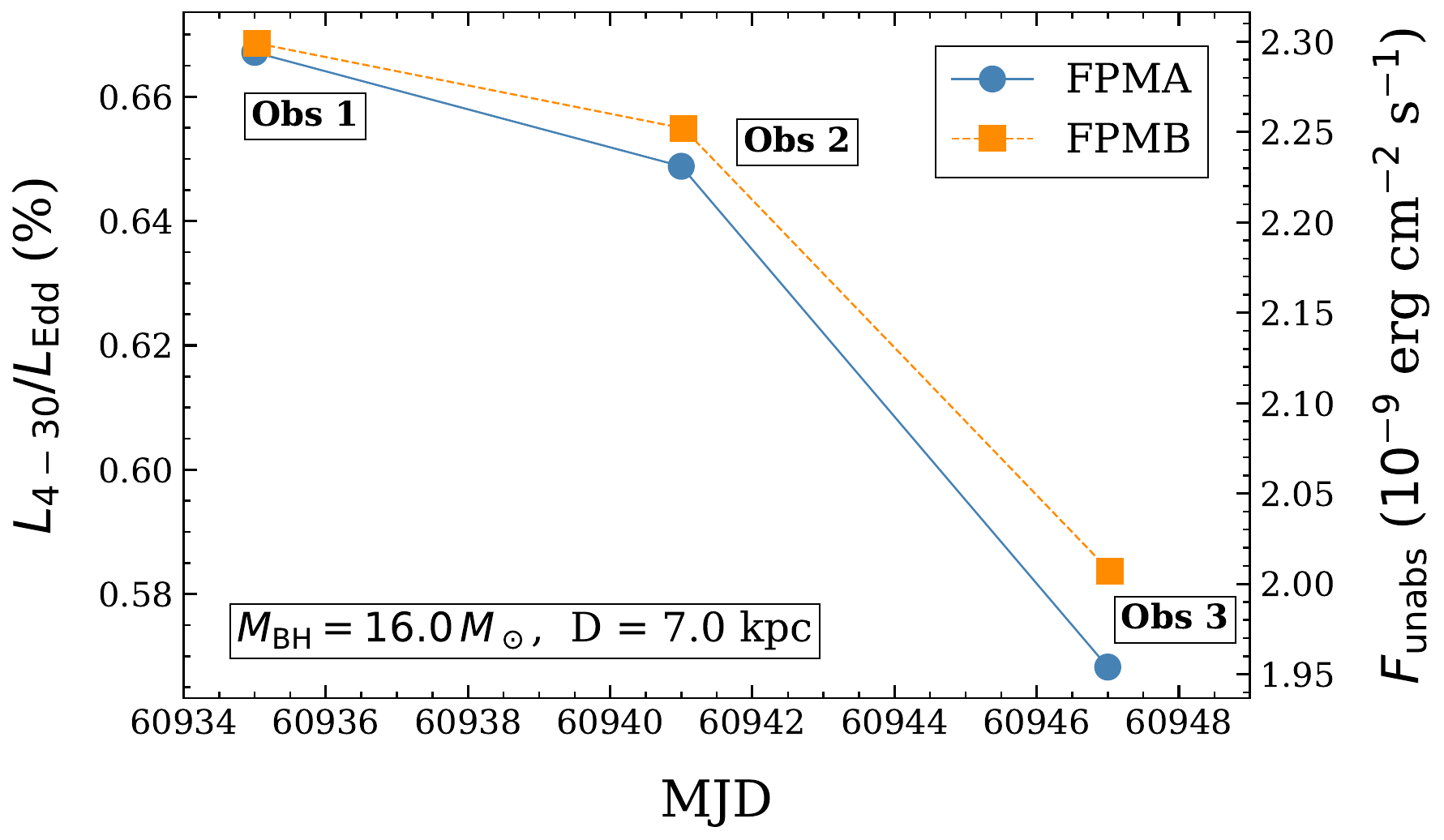}}
        \vspace{-0.2cm}
	\caption{Unabsorbed $4$-$30$\,keV \grs\ luminosity in units of the 
Eddington luminosity $\ledd=2.02\times10^{39}$\,erg\,s$^{-1}$ with the black
hole mass $M=16\,\msun$ \cite{zhao26a} for the three \nustar's observations. 
The steady decline tracks the fading of the 2025 mini-outburst.}
\label{fig:ledd}
\end{figure}

Absorbed and unabsorbed $4$-$30$\,keV fluxes were computed for each 
observation from the best-fit \fpma\ model using \xspec's \texttt{cflux} 
convolution model. To convert these into luminosities, we adopted the distance 
$D=7$\,kpc and BH mass $M=16\,\msun$ as reported by Zhao et 
al.\ \cite{zhao26a}. The Eddington luminosity for this mass is
\begin{equation}
\ledd = 1.26\times10^{38}\left(\frac{M}{\msun}\right)\ {\rm erg\,s^{-1}} = 2.02\times10^{39}\ {\rm erg\,s^{-1}}.
\end{equation}
We computed the isotropic $4$-$30$\,keV luminosity simply as 
$L=4\pi D^2 F_{\rm unabs}$. Here, $F_{\rm unabs}$ is the flux that 
the source would produce in the absence of interstellar absorption. It is 
obtained from the same best-fit model, with the \tbabs\ column set to zero, 
holding every other parameter fixed at its best-fit value. This is the 
physically relevant quantity for computing the source's intrinsic luminosity. 
$F_{\rm abs}$, by contrast, is just what actually reaches the detector, after 
some of the source's true flux has been absorbed along the way. It is a loss 
that has nothing to do with the source itself. The results are summarized in 
Table~\ref{tab:flux} and 
Figure~\ref{fig:ledd}. The unabsorbed flux and luminosity drop by about 
$15\%$ between Obs.\,1 and Obs.\,3. Throughout all three observations, the 
source stays at a persistently low Eddington fraction, 
$L/\ledd\simeq0.57$-$0.67\%$. That is inside the regime where the soft state 
is expected to hold, sources typically don't flip back to the hard state until 
the luminosity drops to around $L/\ledd\gtrsim1$-$2\%$, based on the 
transition luminosities discussed by Maccarone and by Kalemci et 
al.\ \cite{maccarone03a,kalemci13a}. In other words, none of our three epochs 
is anywhere close to the luminosity where we would expect the source to flip 
back into the hard state.

One thing worth stressing: the luminosities we quote only cover the \nustar\ 
energy band, not the source's total (bolometric) output. In the soft state, 
the spectrum is dominated by the disk, which has a temperature of only 
$T_{\rm in}\sim1$\,keV. A blackbody of this temperature actually 
radiates most of its light below $4$\,keV - outside the energy range we fit 
with \nustar. So a great portion of the source's true total brightness is 
simply invisible to our measurement. This means the true, bolometric 
$L/\ledd$ is higher than the values quoted here, reasonably by a factor of 
several orders.
\begin{table}[!h]
	\centering
	\caption{\nustar's estimated flux and luminosity (\fpma, $4$-$30$\,keV) 
of \grs.}\vspace{2pt}
	\label{tab:flux}
	\begin{tabular}{@{}lccc@{}}
\toprule								
& Obs.\,1 & Obs.\,2 & Obs.\,3 \\
\midrule
$F_{\rm abs}$ ($10^{-9}$ erg cm$^{-2}$ s$^{-1}$) & $2.093$ & $2.037$ & $1.777$ \\
$F_{\rm unabs}$ ($10^{-9}$ erg cm$^{-2}$ s$^{-1}$) & $2.294$ & $2.231$ & $1.954$ \\
$L_{\rm unabs}$ ($10^{37}$ erg s$^{-1}$) & $1.345$ & $1.308$ & $1.146$ \\
$L/\ledd$ (\%) & $0.667$ & $0.649$ & $0.568$ \\
	\bottomrule
	\end{tabular}
\end{table}

\subsection{Inner Disk Radius}
\label{sec:rin}

The \diskbb\ normalization is related to an apparent inner disk radius 
through the relation
\begin{equation}
R_{\rm app} = D_{10}\sqrt{\frac{N_{\diskbb}}{\cos i}}\ {\rm km},
\end{equation}
where $D_{10}=D/10\,{\rm kpc}$. 
Using $D=7$\,kpc and the inclination $i=54^\circ$ reported by Zhao et al.\ 
\cite{zhao26a}, the disk normalization, which stayed constant at 
$N_{\diskbb}\approx650$-$653$ across all three observations 
(Table~\ref{tab:specpar}), gives an apparent inner radius of 
$R_{\rm app}\approx23.3$\,km in each case. This apparent radius isn't the true 
physical radius. It needs two standard corrections. The first accounts for 
the fact that the disk's true, local emission isn't a perfect blackbody 
emission. Photons escape from slightly different depths than a blackbody would 
suggest, which ``hardens" the observed spectrum relative to the true effective 
temperature. We correct for this using the color-correction factor 
$f_{\rm col}=1.7$, following Shimura \& Takahara \cite{shimuratakahara95a}. 
The second correction deals with a different simplification. The standard disk 
model assumes a simplified, zero-torque inner boundary. But a fully general 
relativistic disk doesn't quite behave this way. To correct for this 
difference, we use the factor $\xi=0.412$, as calibrated by Kubota et al.\ 
\cite{kubota98a}. Thus, the true and physical inner radii can be calculated
as
\begin{align}
R_{\rm true} & = f_{\rm col}^2\,R_{\rm app},\\[5pt] 					
%
R_{\rm phys} & = \xi\, R_{\rm true}.
\end{align}

%
Applying both corrections, we get a physical inner radius of 
$R_{\rm phys}=27.7$-$27.8$\,km in all three observations. We chose to use the 
same fixed color-correction factor as Shimura \& Takahara 
\cite{shimuratakahara95a}, rather than switching to newer 
disk-atmosphere-model-dependent corrections, simply to stay consistent with 
the value Kubota et al.\ \cite{kubota98a} used when deriving $\xi$ in the 
first place. It is worth noting that $f_{\rm col}$ itself isn't fixed in 
nature; it can vary by several tens of percent depending on the model and 
the local accretion rate. We treat this as an extra systematic uncertainty of 
roughly $10$-$20\%$ on $R_{\rm phys}$, on top of the purely statistical errors 
that come from propagating the uncertainty on $N_{\diskbb}$.

For comparison, we consider the ISCO radius for a Kerr black hole of spin $a$ 
as given by
\begin{equation}
	\risco = \rg\left[3+Z_2-\sqrt{(3-Z_1)(3+Z_1+2Z_2)}\right],
\end{equation}
where
%
	$Z_1=1+(1-a^2)^{1/3}\left[(1+a)^{1/3}+(1-a)^{1/3}\right]$,
	$Z_2=\sqrt{3a^2+Z_1^2}$ and $\rg=GM/c^2=1.4766\,(M/\msun)$\,km,
%
following the classic derivation of Bardeen, Press \& Teukolsky 
\cite{bardeen72a} for the Kerr metric \cite{kerr63a}. For $M=16\,\msun$ and 
$a=0.994$ \cite{zhao26a}, $\rg=23.63$\,km and $\risco=1.235\,\rg=29.18$\,km. 
The derived $R_{\rm phys}/\risco\approx0.95$ in all three observations is 
therefore consistent, within the systematic uncertainty of the 
color-correction procedure, with a disk extending to the ISCO throughout the 
monitored decline.
\begin{table}[!h]
	\centering
	\caption{Inner disk radius (\fpma) of \grs, calculated with 
$f_{\rm col}=1.7$, $\xi=0.412$, $\risco=29.18$\,km.}\vspace{2pt}
	\label{tab:rin}
	\begin{tabular}{@{}lccc@{}}
\toprule
 & Obs.\,1 & Obs.\,2 & Obs.\,3 \\
\midrule
$R_{\rm app}$ (km) & $23.29$  & $23.30$ & $23.36$ \\
$R_{\rm true}$ (km) & $67.31$ & $67.34$ & $67.51$ \\
$R_{\rm phys}$ (km) & $27.73\pm0.12$ & $27.73\pm0.10$ & $27.77\pm0.13$ \\
$R_{\rm phys}/\risco$ & $0.950$ & $0.951$ & $0.952$ \\
\bottomrule
	\end{tabular}
\end{table}

\subsection{Parameter Evolution Across the Outburst Decline}
\label{sec:paramevol}

\begin{figure}[!h]
	\centerline{
	\includegraphics[scale=0.3]{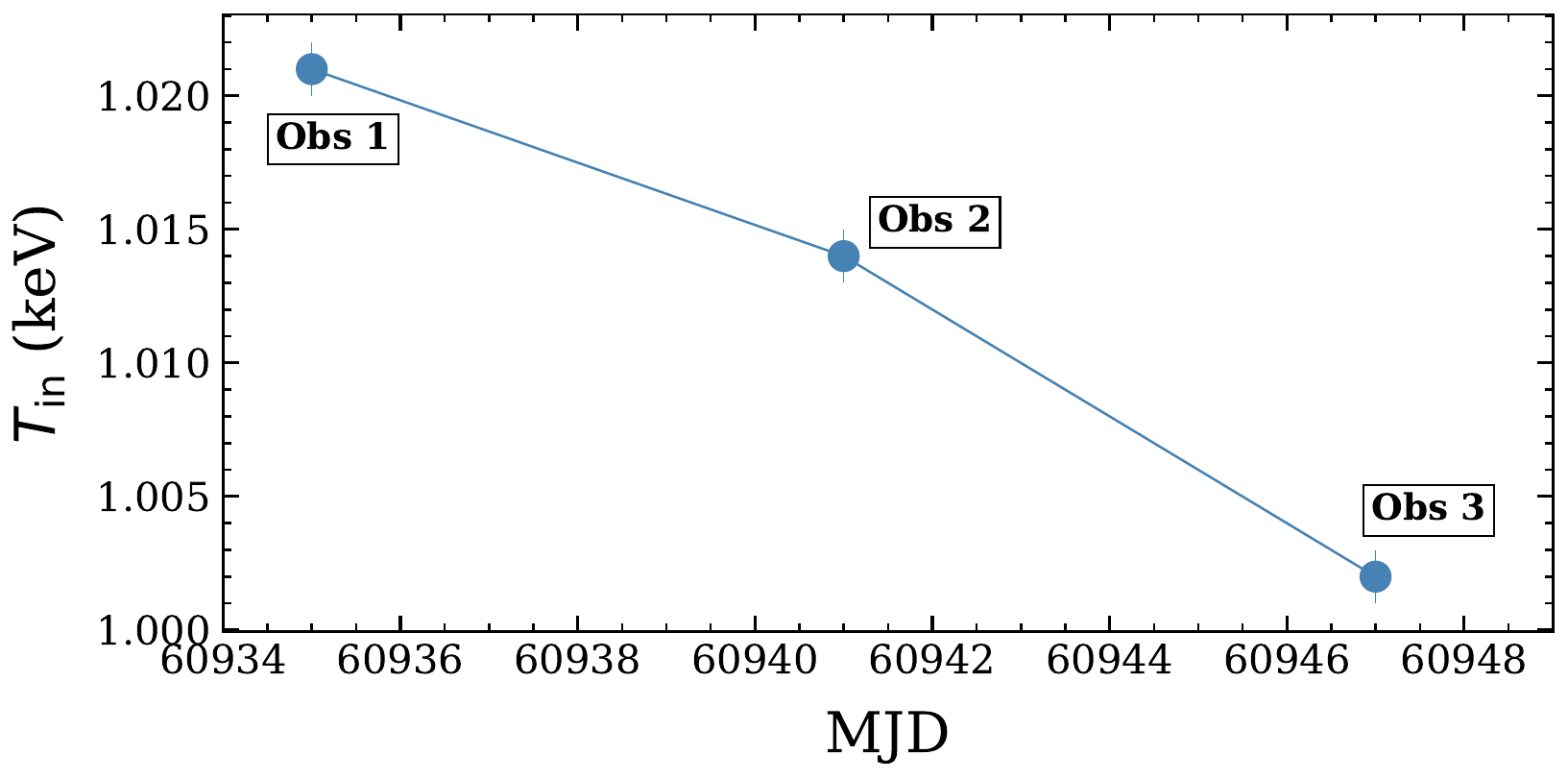}\hspace{0.3cm}
	\includegraphics[scale=0.3]{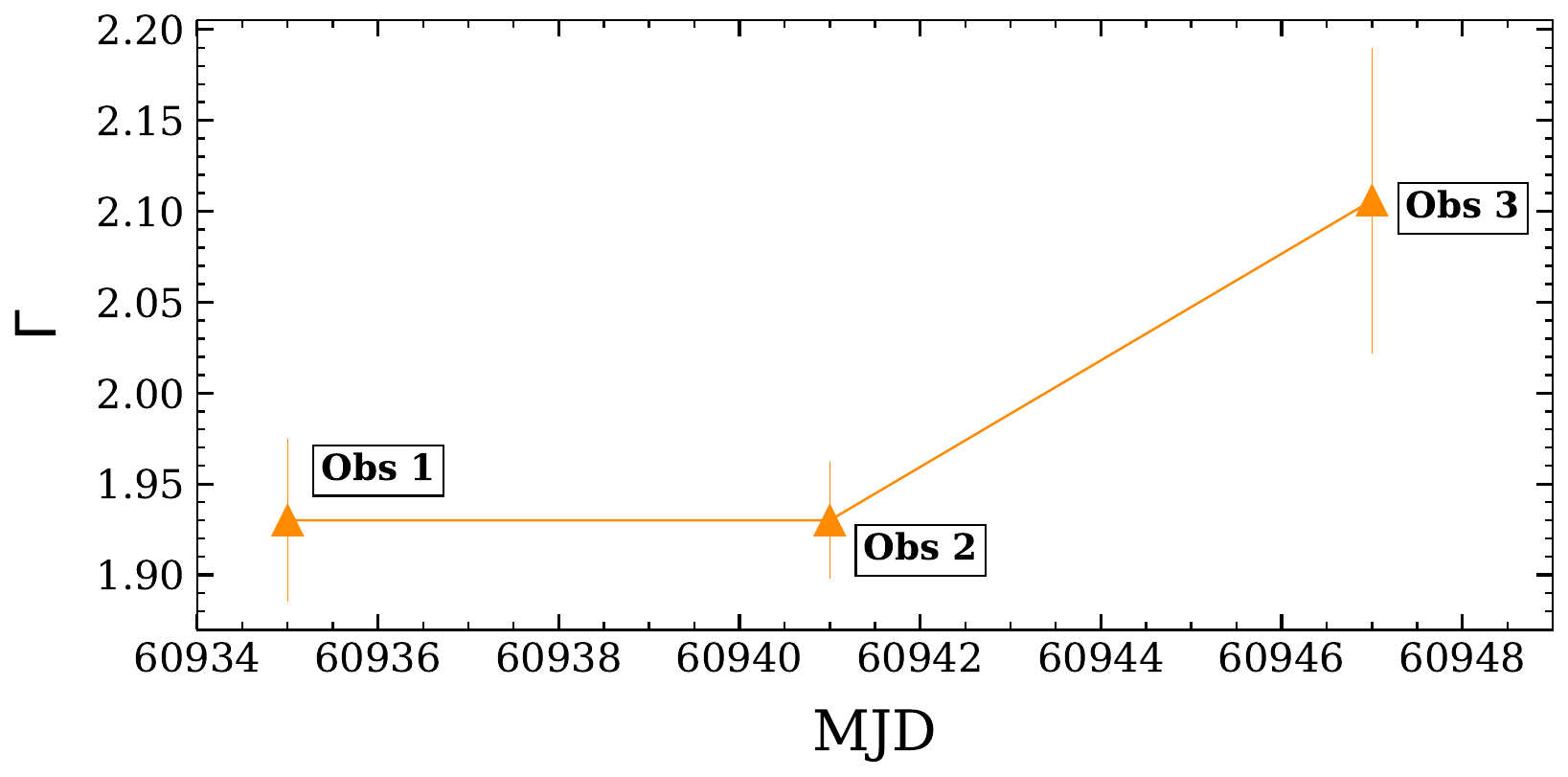}}\vspace{0.3cm}
        \centerline{
	\includegraphics[scale=0.3]{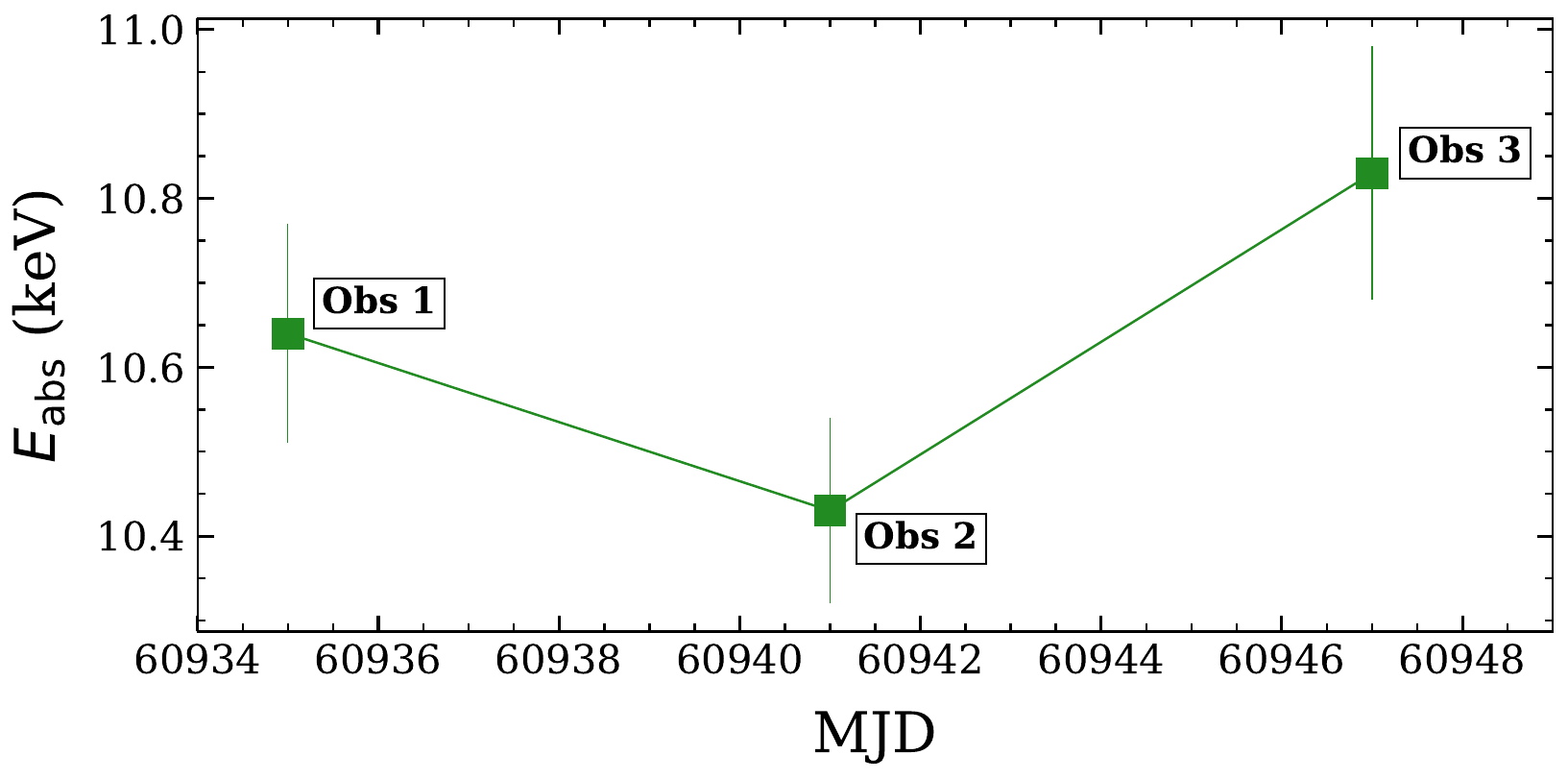}\hspace{0.3cm}
	\includegraphics[scale=0.3]{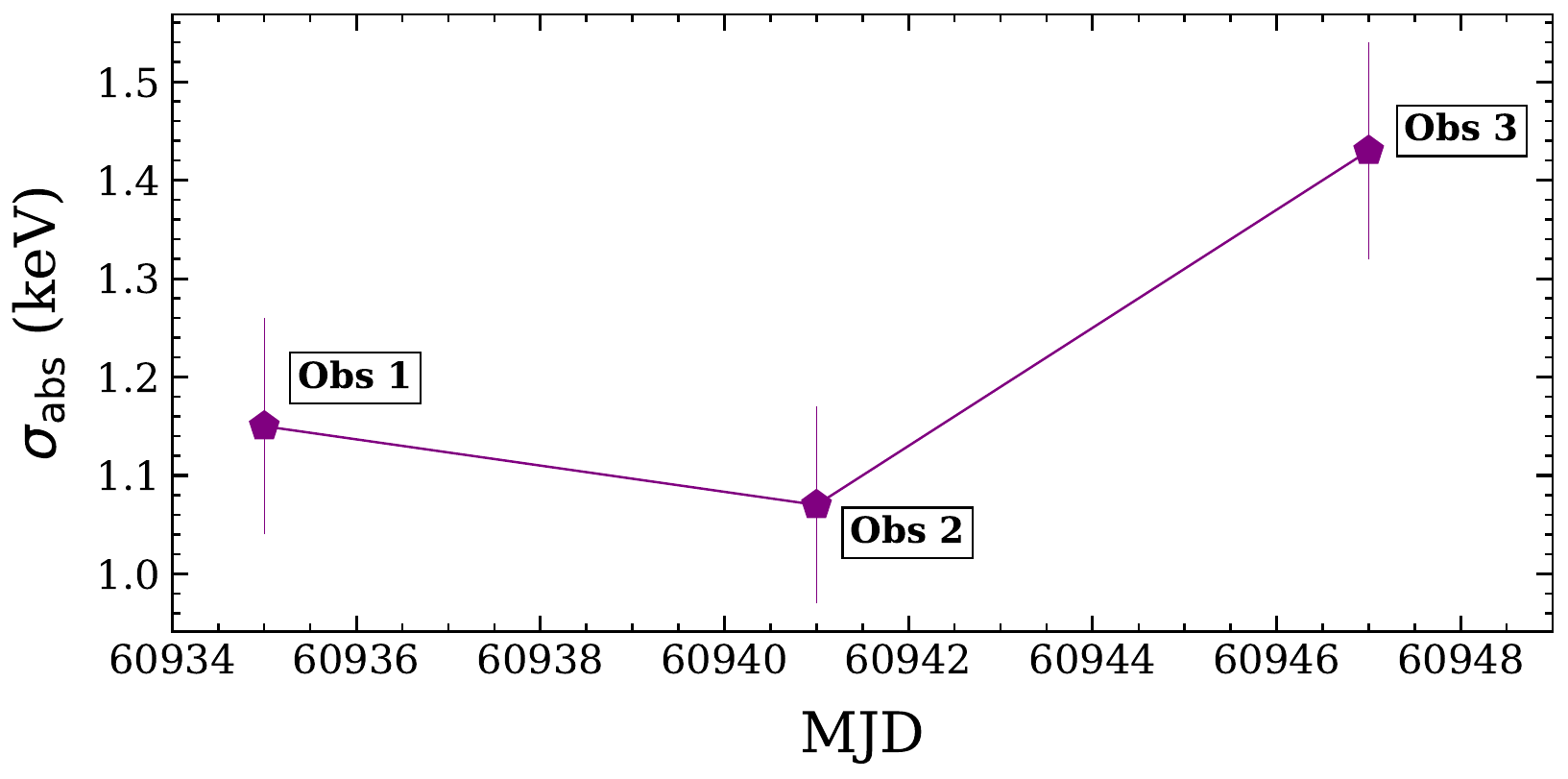}}\vspace{0.3cm}
        \centerline{
	\includegraphics[scale=0.3]{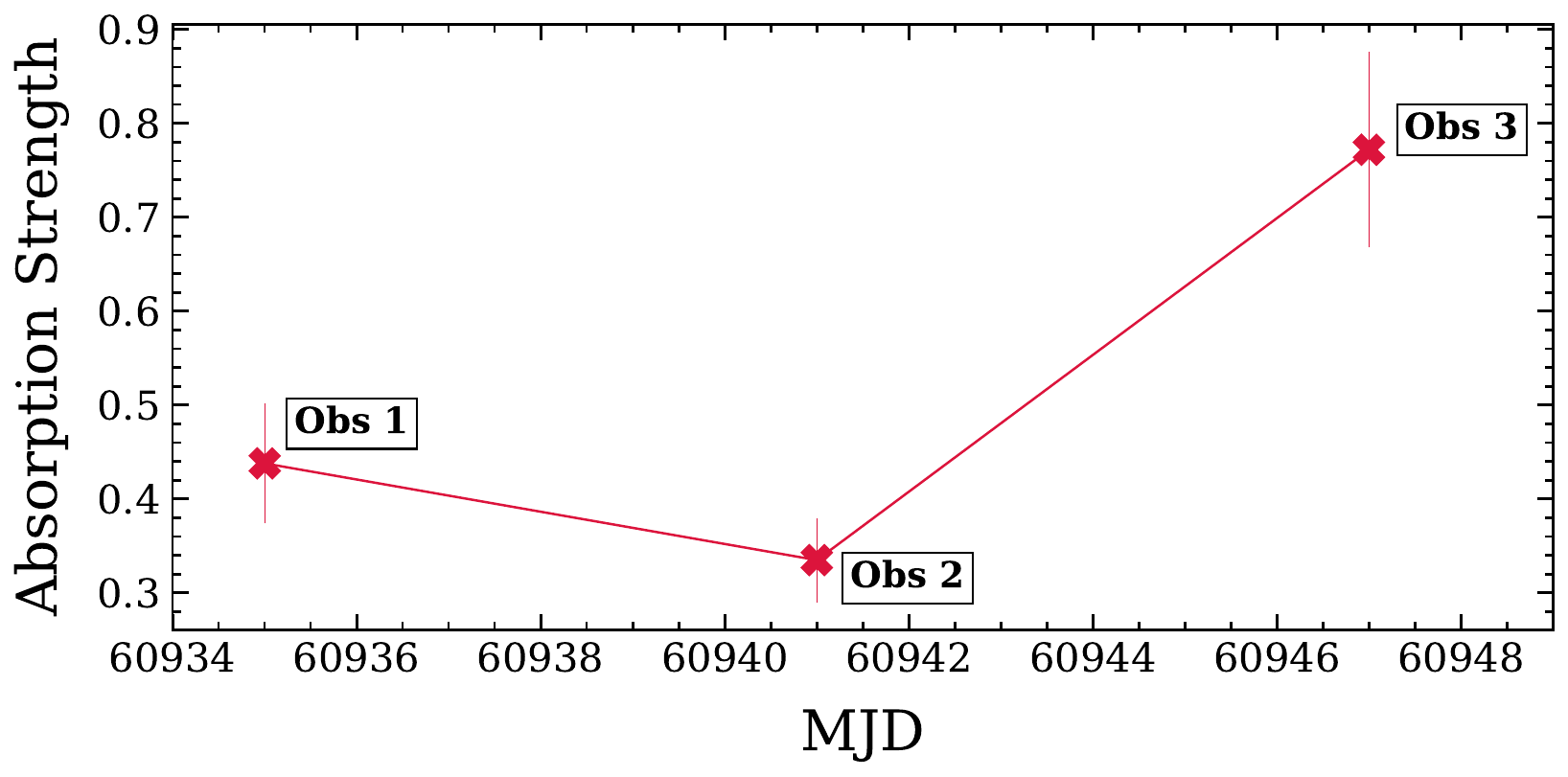}}
        \vspace{-0.2cm}
	\caption{Evolution of $T_{\rm in}$, $\Gamma$ and the gabs 
strength and centroid energy across the three \nustar\ observations of \grs.}
	\label{fig:paramevol}
\end{figure}

Table~\ref{tab:specpar} and Figure~\ref{fig:paramevol} show that the free 
parameters of our joint spectral model do not all evolve in the same way 
across the $12$-day span of the three observations, and hence it is worth 
examining each in turn.

The cross-normalization constant $C_{\fpmb}$ increases slowly but 
monotonically, from $1.002\pm0.003$ in Obs.\,1 to 
$1.010\pm0.003$ in Obs.\,2 and $1.027^{+0.003}_{-0.002}$ in Obs.\,3. All three 
values remain within the typical range of \fpma/\fpmb\ cross-calibration 
residuals reported by Madsen et al.\ \cite{madsen15a} and by other \nustar\ 
studies of bright point sources \cite{harrison13a,miller15a,fuerst16a}. So we 
do not regard this trend as astrophysically significant. It most plausibly 
reflects a slowly time-variable, low-level calibration residual between the 
two focal-plane modules rather than any change intrinsic to \grs\ itself.

The disk temperature $T_{\rm in}$ declines steadily with high statistical 
significance across all three epochs, from $1.021\pm0.001$\,keV (Obs.\,1) 
through $1.014\pm0.002$\,keV (Obs.\,2) to $1.002^{+0.003}_{-0.002}$\,keV 
(Obs.\,3), a $\sim2\%$ decline. Given how precisely $T_{\rm in}$ is 
constrained in each fit, to well under a percent, this decline is highly 
significant. For a standard disk, the luminosity 
$L_{\rm disk}\propto R_{\rm in}^2 T_{\rm in}^4$. Since the emitting area 
stays essentially constant, this decline in $T_{\rm in}$ alone accounts for 
nearly all of the drop in luminosity we see (Table~\ref{tab:flux}). This is 
exactly what we would expect for a disk that is cooling as the mass-accretion 
rate declines, while its inner edge stays pinned at a fixed radius.

The disk normalization $N_{\diskbb}$ and hence the derived inner disk radius 
is by contrast remarkably stable: $650.5\pm8.9$, $650.8\pm7.8$ and 
$652.5^{+9.7}_{-10.4}$ in Obs.\,1-3, respectively. It is formally consistent 
with a constant value within the statistical errors (Section~\ref{sec:rin}). 
Together with the declining $T_{\rm in}$, this stability is the key piece of 
spectroscopic evidence in this paper for a disk that stays anchored at a 
fixed inner radius throughout the decline (Section~\ref{sec:rin}). We discuss 
the physical interpretation of this result in Section~\ref{sec:results}.

The Comptonized tail photon index $\Gamma$ and the parameters of the \gabsm\ 
component evolve in a distinctly non-monotonic, two-phase pattern. Obs.\,1 
and Obs.\,2, separated by six days, are close to indistinguishable in 
$\Gamma$ (both $1.930$, with overlapping $90\%$ confidence intervals), in the 
\gabsm\ centroid energy ($10.639^{+0.235}_{-0.216}$ and 
$10.427^{+0.181}_{-0.172}$\,keV, respectively) and width 
($\sigma=1.146^{+0.180}_{-0.165}$ and $1.067^{+0.147}_{-0.133}$\,keV). The 
\gabsm\ strength is marginally lower in Obs.\,2 ($0.335^{+0.074}_{-0.063}$) 
than in Obs.\,1 ($0.438^{+0.118}_{-0.095}$), though the two are consistent 
within their $90\%$ uncertainties. By Obs.\,3, a further six days later, this 
comparatively stable configuration gives way to something different. The 
photon index $\Gamma$ steepens to $2.106^{+0.151}_{-0.135}$. At the same time, 
the \gabsm\ component broadens ($\sigma=1.426^{+0.207}_{-0.183}$\,keV) and 
strengthens sharply (strength $=0.772^{+0.216}_{-0.158}$, more than double the 
Obs.\,2 value), while its centroid shifts to the highest energy seen across 
the three epochs ($10.828^{+0.286}_{-0.240}$\,keV). The normalization of the 
\nthcomp\ component itself falls by a factor of $\sim2$-$2.5$ between 
Obs.\,2 and Obs.\,3 ($1.723^{+0.19}_{-0.17}\times10^{-2}$ to 
$0.678^{+0.23}_{-0.16}\times10^{-2}$), even though $\Gamma$ steepens. It 
indicates that the Comptonized tail itself is both weakening and softening in 
the final observation, at the same time as the broad curvature feature (that 
we associated with reflection) grows substantially stronger. We regard this 
as evidence for two distinct phases within the monitored decline - a 
comparatively steady early phase (Obs.\,1-2) followed by a more abrupt change 
in the coronal and reflection properties by Obs.\,3. It's worth noting that 
this second phase lines up exactly with the epoch that \textit{IXPE} 
polarimetry independently confirmed to be dominated by a strong, 
returning-radiation reflection signal from a near-extremal-spin black hole 
\cite{zhao26a}. We discuss this coincidence further in 
Section~\ref{sec:results}.

Two cautions apply to this parameter-by-parameter picture. First, the \gabsm\ 
parameters are not truly independent of each other. The centroid, width and 
strength all jointly shape the broad curvature they are fit to reproduce. So 
some of the apparent structure we see in their individual trends is better 
thought of as describing the overall shape of that curvature, rather than as 
three separate physical measurements. Second, the $\Gamma$ shapes the 
continuum across the entire fitted range, $4$-$30$\,keV. On the other hand, 
the \gabsm\ component only reproduces curvature over a narrower stretch, 
roughly $7$-$14$\,keV, based on its fitted centroid and width. These two 
ranges overlap. Because of that overlap, the fit has some room for one to 
compensate for the other: a softer intrinsic continuum in that overlapping 
region can partly mimic the effect of a stronger, broader \gabsm\ component 
and vice versa. 

\begin{figure}[!h]
\centerline{
\includegraphics[scale=0.4]{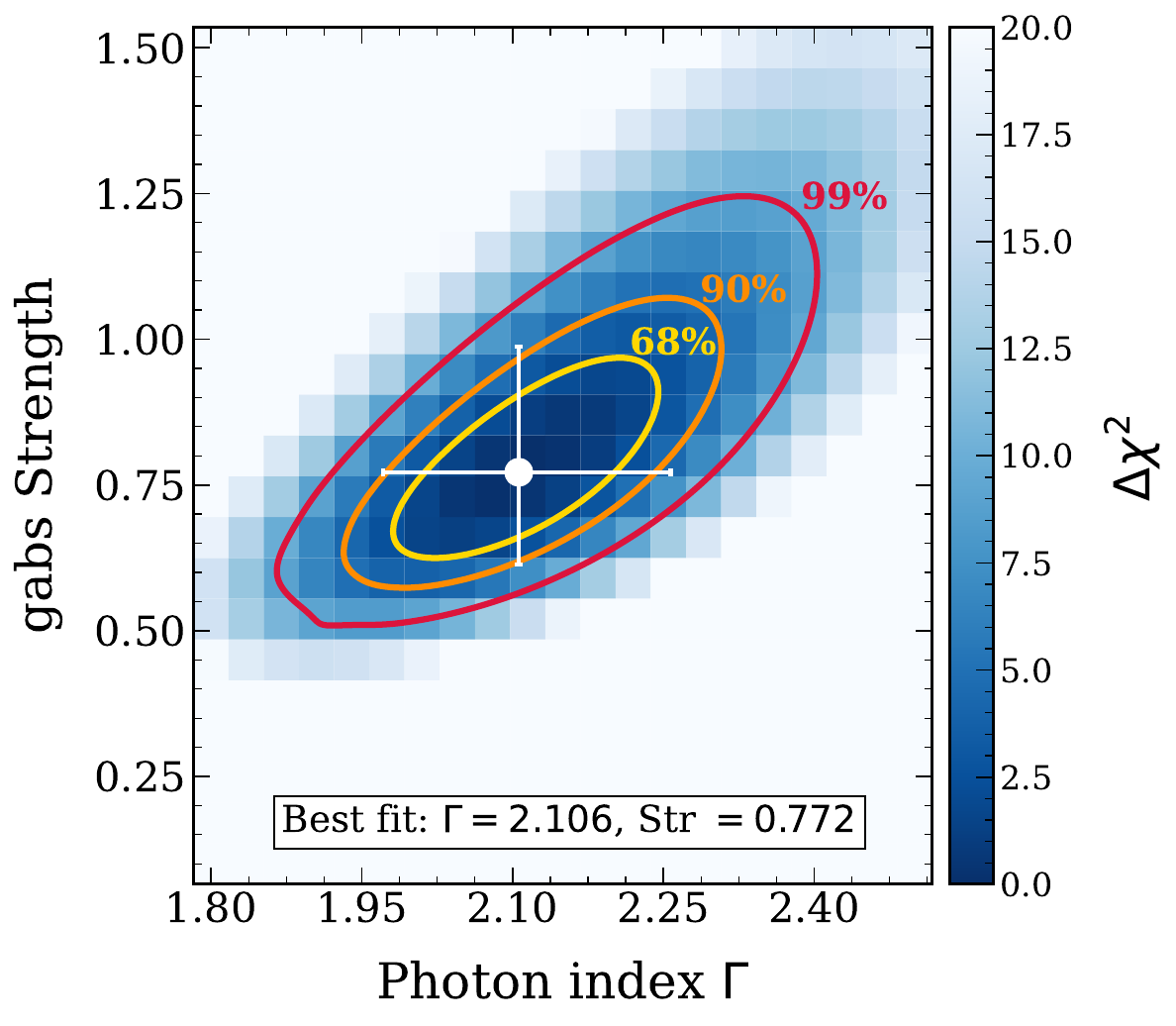}}
\vspace{-0.2cm}
\caption{Two-dimensional $\chi^2$ confidence contours ($68\%$, $90\%$ and 
$99\%$; $\Delta\chi^2=2.30$, $4.61$ and $9.21$) from a \texttt{steppar} grid 
in the photon index $\Gamma$ and \gabsm\ strength, for the joint \fpma+\fpmb\ 
fit to Obs.\,3. The contours form a single, closed region around the best fit 
($\Gamma=2.106$, Strength$=0.772$; white point with $1\sigma$ error bars), 
elongated diagonally that reflects a mild positive correlation between the 
two parameters.}
\label{fig:steppar}
\end{figure}
The fit could, in principle, favor one over the other without a huge cost in 
$\chi^2$. To check whether this degeneracy was actually a problem, we computed 
a two-dimensional \texttt{steppar} grid in $\Gamma$ and the \gabsm\ strength 
for Obs.\,3 (Figure~\ref{fig:steppar}). The resulting $\chi^2$ landscape shows 
a single, well-defined minimum, coincident with our adopted best fit 
($\Gamma=2.106$, strength$=0.772$), with nested $68\%$, $90\%$ and $99\%$ 
confidence contours. There is no sign of a separate, competing solution. 
The contours are noticeably elongated diagonally, indicating a mild positive 
correlation between $\Gamma$ and the \gabsm\ strength. This means that if we 
make the continuum a little softer while also making the \gabsm\ component a 
little stronger, the fit barely gets any worse. The same is true the other 
way around, with a slightly harder continuum and a slightly weaker \gabsm\ 
component. So the two parameters are not completely independent of each other. 
This correlation is modest enough, however, that it does not change our 
conclusions qualitatively. To fully separate these two effects with 
confidence, we would need a self-consistent reflection model rather than the 
simplified empirical approach we use here.

\section{Results and Discussion}
\label{sec:results}

Every independent diagnostic examined in this work - the $3$-$79$\,keV count 
rate, the hardness ratio, the fractional variability, the rms spectrum, the 
absorbed and unabsorbed flux and the derived luminosity-declines from Obs.\,1 
to Obs.\,3, gives a consistent picture of a soft-state black hole binary 
fading over the $12$-day span of our \nustar\ monitoring 
(Tables~\ref{tab:timing}-\ref{tab:flux}). The disk temperature $T_{\rm in}$ 
declines steadily from $1.021\pm0.001$\,keV to $1.002^{+0.003}_{-0.002}$\,keV, 
while the disk normalization and hence the derived physical inner radius 
remains constant to within $0.3\%$ across all three epochs 
(Table~\ref{tab:rin}). This is the expected signature of 
a standard disk that stays anchored at the ISCO while cooling in response to 
a declining mass-accretion rate. A radially receding, truncated disk would 
look different because, as the source faded, its emitting area and hence 
its normalization would be expected to shrink along with it. That is not what 
we see here. This same diagnostic, a constant $N_{\diskbb}$ alongside a 
declining $T_{\rm in}$, was used to argue for a stable, ISCO-anchored disk 
in other black hole transients monitored across multiple epochs of a soft 
state \cite{mcclintock14a,steiner10a}.

The spectral evolution of the Comptonized tail and the broad curvature feature 
are less monotonic (Section~\ref{sec:paramevol}). Obs.\,1 and Obs.\,2 are 
close to indistinguishable in $\Gamma$ (both $1.930$) and in the \gabsm\ 
parameters (Table~\ref{tab:specpar}, Figure~\ref{fig:paramevol}). They also 
occupy the same locus in the HID (Figure~\ref{fig:hid}). Obs.\,3, by contrast, 
shows a steeper photon index ($\Gamma=2.106^{+0.151}_{-0.135}$) and a broader, 
stronger curvature feature (centroid $10.83$\,keV, $\sigma=1.43$\,keV, 
strength $0.77$, versus $\sigma\lesssim1.15$\,keV and strength $\lesssim0.44$ 
in Obs.\,1-2). We interpret this as evidence for two phases within the 
monitored decline: a comparatively stable early phase (Obs.\,1-2, separated 
by six days) followed by a further, more pronounced softening and 
strengthening/broadening of the reflection-like feature by Obs.\,3 (a further 
six days later). It is dominated by a strong disk-reflection signal from 
gravitationally returning radiation around a near-extremal-spin BH 
\cite{zhao26a}. The significance of the \gabsm\ component itself increases 
monotonically across the three epochs ($11.7\sigma \rightarrow 12.4\sigma 
\rightarrow 15.0\sigma$; Table~\ref{tab:ftest}). This is consistent with this 
picture, since a stronger, broader feature is intrinsically easier to 
detect at fixed exposure.

The timing properties reinforce this picture of a settled, disk-dominated 
soft state rather than a source undergoing dramatic coronal restructuring. The 
absence of any significant broadband noise, QPO or soft-hard time lag in any 
of the three observations (Sections~\ref{sec:timing}-\ref{sec:rmsspec}) is 
consistent with a weakly variable, disk-dominated soft state and with the low 
fractional variability and flat rms spectrum measured directly from the light 
curves. On its own, this null result can not tell us which of two things is 
going on. Either the corona is genuinely compact and weak, similar to what 
Kara et al.\ \cite{kara19a} found shrinking on timescales of days during the 
hard-to-soft transition of MAXI~J1820+070, later characterized further by 
Buisson et al.\ \cite{buisson19a}. Or, alternatively, the corona's variability 
is simply too faint for us to pick up at the count rates and exposures 
available here. Our data can not distinguish between these two possibilities. 
We do note something interesting, though. The moderate but real changes 
in $\Gamma$ and the \gabsm\ parameters between Obs.\,2 and Obs.\,3 
(Section~\ref{sec:paramevol}) happen without any matching large change in 
$\fvar$ or the rms spectrum. This suggests that whatever changes in coronal 
or reflection geometry are driving the spectral evolution are not accompanied 
by an equally dramatic change in the amplitude of the fast aperiodic 
variability. This same combination - a real spectral shift without a matching 
change in variability - has also been seen during the compact-corona phases of 
other well-monitored black hole transients, such as GX~339-4 
\cite{fuerst15a,parker16a} and MAXI~J1820+070 \cite{buisson19a,kara19a}.

Our derived inner-disk radius and its stability at 
$R_{\rm phys}/\risco\approx0.95$ throughout the decline is consistent with 
what Zhao et al.\ \cite{zhao26a} found. This gives us an independent, purely 
spectral confirmation of the near-extremal spin and high-inclination geometry 
that Zhao et al.\ \cite{zhao26a} inferred from simultaneous X-ray polarimetry 
of Obs.\,3. It is worth comparing this to the earlier \nustar\ observations 
of the source. Our present soft-state epochs look quite different from the 
studies done by Miller et al.\ \cite{miller15a} and F\"urst et al.\ 
\cite{fuerst16a}. The bright hard state analyzed by Miller et al.\ 
\cite{miller15a} showed almost no disk contribution at all, with the spectrum 
dominated by a broad, relativistic Fe\,K$\alpha$ line, modeled using the same 
family of relativistic reflection formalisms applied across the wider BH XRB 
population \cite{fabian89a,laor91a,garcia14a,dauser13a}. The very faint hard 
state analyzed by F\"urst et al.\ \cite{fuerst16a} was different again. The 
disk was still only a minor contributor. But the continuum was unusually hard, 
with evidence of a truncated disk, similar to the truncated-disk phases seen 
in other black hole transients as they approach quiescence 
\cite{narayan95a,doneGK07}. Now, there are four \nustar-era observations that 
exist for \grs: the 2014/2015 bright hard state \cite{miller15a}, the 
2014/2015 faint hard state \cite{fuerst16a}, and the two soft-state phases of 
the 2025 mini-outburst covered here and by Zhao et al.\ \cite{zhao26a}. Across 
these four epochs, the source's inner accretion geometry clearly changes 
depending on which state it is in. The disk only reaches the ISCO once the 
source has entered the soft state and gets truncated in the hard state. This 
pattern has been established by both hard-state and soft-state studies like 
\cite{mcclintock14a,doneGK07,remillard06a}.

\section{Summary and Conclusion}
\label{sec:conclusion}

We have presented a spectral and timing analysis of three \nustar\ 
observations of the black hole X-ray binary \grs\ obtained during the decline 
of its 2025 mini-outburst. Across all three epochs, the source is in a soft, 
disk-dominated state, with a declining count rate, a low and only mildly 
evolving hardness ratio, a fractional variability that decreases from $3.2\%$ 
to $1.1\%$ and a comparatively flat, low-amplitude rms spectrum. No periodic 
or quasi-periodic signal is detected in any observation. A discrete 
correlation function analysis reveals no significant soft-hard time lag. A 
joint \fpma+\fpmb\ spectral model consisting of an absorbed disk-blackbody 
plus thermal-Comptonization continuum, together with a broad curvature feature 
near $10$-$12$\,keV (that we interpret as an empirical proxy for a 
Compton-reflection/returning-radiation component) provides a good description 
of the $4$-$30$\,keV spectra in all three observations 
($\chi^2/{\rm dof}=1.04$-$1.15$). The curvature feature itself shows up 
strongly at $11.7$-$15.0\sigma$ in a standard $F$-test. This test is known to 
have calibration issues. But even taking those into account, we think the 
preference for this extra component holds up. The inner-disk temperature 
cools steadily from $1.021$ to $1.002$\,keV while the disk normalization and 
hence the derived physical inner radius, $R_{\rm phys}\approx27.7$\,km 
$\approx0.95\,\risco$, remains constant to within $0.3\%$. This is consistent 
with a disk anchored at the ISCO throughout the monitored decline rather than 
one that is radially receding as the source fades. The unabsorbed 
$4$-$30$\,keV luminosity declines by $\sim15\%$ across the three epochs 
from $1.35\times10^{37}$ to $1.15\times10^{37}$\,erg\,s$^{-1}$ ($0.67\%$ to 
$0.57\%\,\ledd$). This is consistent with the fading of the mini-outburst. 
The photon index and the \gabsm\ parameters evolve in a distinctly 
non-monotonic, two-phase pattern: Obs.\,1 and Obs.\,2 are close to 
indistinguishable from one another, while the broad curvature feature 
strengthens and broadens substantially by Obs.\,3. It is coincident with the 
epoch independently confirmed by \textit{IXPE} polarimetry to show a strong 
reflection signal from returning radiation around a near-extremal-spin black 
hole. It suggested the possibility of genuine spectral evolution within the 
soft state on a timescale of days rather than a single, static configuration. 
Taken together, these results push the spectral-timing picture of the 2025 
mini-outburst of \grs\ beyond the single simultaneous \textit{IXPE} epoch. 
They also give an independent, purely spectroscopic check on the near-extremal 
spin and high-inclination geometry that Zhao et al.\ \cite{zhao26a} infer from 
X-ray polarimetry. They fit well with the broader pattern seen across BH X-ray 
binaries - a truncated disk in the hard state versus a disk anchored at the 
ISCO in the soft state. Looking ahead, one thing that would help test these 
conclusions more rigorously: future observations that combine simultaneous 
broadband spectroscopy and polarimetry across a wider span of the outburst 
would let us check whether the disk really does stay at the ISCO throughout 
the soft state.

\section{Acknowledgements}
\label{sec:ack}

This research has made use of data obtained with the \nustar\ mission, a 
project led by the California Institute of Technology, managed by the Jet 
Propulsion Laboratory and funded by NASA. We thank the NASA \heasoft\ and 
\nustar\ Guest Observer program teams for developing and maintaining the 
\heasoft, \texttt{NuSTARDAS} and \xspec\ software packages and for the 
technical support and documentation that made this analysis possible. This 
work has also made use of the \texttt{Stingray} spectral-timing package and 
of NASA's Astrophysics Data System. UDG is also thankful to the 
Inter-University Centre for Astronomy and Astrophysics (IUCAA), Pune, India 
for the Visiting Associateship of the institute.


\end{document}